\documentclass[final,3p,times]{elsarticle}
\usepackage[utf8]{inputenc}
\DeclareFixedFont{\ttb}{T1}{txtt}{bx}{n}{10} % for bold
\DeclareFixedFont{\ttm}{T1}{txtt}{m}{n}{10}  % for normal
\DeclareFixedFont{\ttmsub}{T1}{txtt}{m}{n}{6}  % for normal, subscript size
\DeclareFixedFont{\ttmfoot}{T1}{txtt}{m}{n}{8}  % for normal, footnote size

\usepackage{xcolor}%\usepackage{color}
\definecolor{deepblue}{rgb}{0,0,0.5}
\definecolor{deepred}{rgb}{0.6,0,0}
\definecolor{deepgreen}{rgb}{0,0.5,0}
\definecolor{gray}{rgb}{0.6,0.6,0.6}

\usepackage{listings}
\newcommand\pythonstyle{\lstset{
mathescape,
language=Python,
xleftmargin=0.1\textwidth,
basicstyle=\ttm,
otherkeywords={self},             % Add keywords here
keywordstyle=\ttb\color{deepblue},
emph={MyClass,__init__},          % Custom highlighting
emphstyle=\ttb\color{deepred},    % Custom highlighting style
stringstyle=\color{deepgreen},
showstringspaces=false,           %
commentstyle=\ttm\color{gray},
backgroundcolor=\color{green!10}
}}

\newcommand\pythonstylepseudo{\lstset{
mathescape,
language=Python,
xleftmargin=0.1\textwidth,
basicstyle=\ttm,
otherkeywords={self},             % Add keywords here
keywordstyle=\ttb\color{deepblue},
emph={MyClass,__init__},          % Custom highlighting
emphstyle=\ttb\color{deepred},    % Custom highlighting style
stringstyle=\color{deepgreen},
showstringspaces=false,           %
commentstyle=\ttm\color{gray},
backgroundcolor=\color{red!10}
}}

\newcommand\OutputStyle{\lstset{
  basicstyle=\small\ttfamily,
  numbers=none,
  frame=tblr,
  columns=fullflexible,
  linewidth=0.9\linewidth,
  xleftmargin=0.1\linewidth,
  backgroundcolor=\color{blue!10}
}}

\lstnewenvironment{python}[1][]{\pythonstyle\lstset{#1}}{}
\newcommand\pyin[1]{{\pythonstyle\lstinline!#1!}}
\lstnewenvironment{outsty}[1][]{\OutputStyle\lstset{#1}}{}
\lstnewenvironment{pythonpseudo}[1][]{\pythonstylepseudo\lstset{#1}}{}

\usepackage{graphicx}
\usepackage{amssymb}
\usepackage{amsthm}
\usepackage{hyperref}
\usepackage{nccmath}
\usepackage[normalem]{ulem} % for sout{} command
\newcommand{\e}{e}
\newcommand{\ii}{i}
\newcommand{\mc}{\mathcal}
\newcommand{\mr}{\mathrm}

\newcommand{\ve}{\varepsilon}

\newcommand{\pd}{\partial}
\newcommand{\dif}{\mathrm{d}}
\newcommand\abs[1]{\lvert#1\rvert}

\newcommand\avgs[1]{\langle#1\rangle}
\newcommand\bra[1]{\langle#1\rvert}
\newcommand\ket[1]{\lvert#1\rangle}
\newcommand{\Real}{\operatorname{Re}}
\newcommand{\Imag}{\operatorname{Im}}
\newcommand{\Tr}{\operatorname{Tr}}
\newcommand{\up}{\uparrow}
\newcommand{\down}{\downarrow}
\newcommand{\cd}{c^{\dag}}
\newcommand{\can}{c^{\phantom{\dag}}}
\newcommand{\dd}{d^{\dag}}
\newcommand{\dan}{d^{\phantom{\dag}}}
\newcommand{\shs}[1]{}
\newcommand{\leadqn}{\alpha}
\newcommand{\leadE}[1]{\ve_{#1}}

\journal{Computer Physics Communications}

\begin{document}

\begin{frontmatter}

\title{\texttt{QmeQ} 1.0: An open-source \texttt{Python} package for calculations of transport through quantum dot devices}

\author[a]{Gediminas Kir{\v{s}}anskas} %\corref{author}
\author[c]{Jonas Nyvold Pedersen}
\author[a]{Olov Karlstr{\"{o}}m}
\author[b]{Martin Leijnse}
\author[a]{Andreas Wacker}

%\cortext[author] {Corresponding author.\\\textit{E-mail address:} qmeq.package@gmail.com}
\address[a]{Mathematical Physics and NanoLund, Lund University, Box 118, 22100 Lund, Sweden}
\address[b]{Solid State Physics and NanoLund, Lund University, Box 118, 221 00 Lund, Sweden}
\address[c]{Department of Micro- and Nanotechnology, Technical University of Denmark, DK-2800 Kgs. Lyngby, Denmark}
%\address[d]{Other Address}

\begin{abstract}
\texttt{QmeQ} is an open-source \texttt{Python} package for numerical modeling of transport through quantum dot devices with strong electron-electron interactions using various approximate master equation approaches. The package provides a framework for calculating stationary particle or energy currents driven by differences in chemical potentials or temperatures between the leads which are tunnel coupled to the quantum dots. The electronic structures of the quantum dots are described by their single-particle states and the Coulomb matrix elements between the states. When transport is treated perturbatively to lowest order in the tunneling couplings, the possible approaches are \textit{Pauli} (classical), \textit{first-order Redfield}, and \textit{first-order von Neumann} master equations, and a particular form of the \textit{Lindblad} equation. When all processes involving two-particle excitations in the leads are of interest, the \textit{second-order von Neumann} approach can be applied. All these approaches are implemented in \texttt{QmeQ}. We here give an overview of the basic structure of the package, give examples of transport calculations, and outline the range of applicability of the different approximate approaches.
\end{abstract}

\begin{keyword}
Open quantum systems, Quantum dots, Anderson-type model, Coulomb blockade, \texttt{Python}
\end{keyword}

\end{frontmatter}

\noindent \textbf{Program summary}

\begin{small}
\noindent\emph{Program Title:} \texttt{QmeQ} \\
\emph{Licensing provisions:} BSD 2-Clause \\
\emph{Programming language:} \texttt{Python} \\
\emph{External libraries:} \texttt{NumPy}, \texttt{SciPy}, \texttt{Cython} \\
%\emph{Supplementary material:} \\
\emph{Nature of problem:} Calculation of stationary state currents through quantum dots tunnel coupled to leads.\\
\emph{Solution method:} Exact diagonalisation of the quantum dot Hamiltonian for a given set of single particle states and Coulomb matrix elements. Numerical solution of the stationary-state master equation for a given approximate approach. \\
\emph{Restrictions: } Depending on the approximate approach the temperature needs to be sufficiently large compared to the coupling strength for the approach to be valid. \\
\end{small}

\section{\label{sec:intro}Introduction}

Quantum dot devices are usually made of a nanostructure or a molecule coupled to leads \cite{ChakrabortyBook1999,SohnBook1997,YuPRL2004}. A vast amount of experiments have been performed using quantum dots, where with bias spectroscopy \cite{HansonRMP2007} or thermoelectric measurements \cite{SvilansPHE2016}, it is possible to obtain the energy level structure of the quantum dot, which is important for understanding and predicting the device behavior. Such a device is also a useful tool for studying the fundamental physics of open-quantum systems in non-equilibrium, where the leads play the role of an environment \cite{BreuerBook2006}. Modeling transport through such nanoscale systems is a complicated task, especially when strong Coulomb interaction is present in the quantum dots, as it can lead to phenomena like Coulomb blockade \cite{GrabertBook1992,DeFranceschiPRL2001},
%Coulomb blockade with cotunneling \cite{DeFranceschiPRL2001},
Kondo effect \cite{GoldhaberNature1998,CronenwettScience1998,PustilnikJPhysCondensMat2004}, or Pauli blockade \cite{OnoScience2002}, just to mention a few. Quantum transport can be calculated using different theoretical methods like scattering-states numerical-renormalization group \cite{AndersPRL2008},
non-equilibrium Green's functions \cite{MeirPRL1992,RyndykBook2009}, and master equation based approaches \cite{BreuerBook2006, NazarovPHB1993, GurvitzPRB1996, KonigPRB1996, PedersenPRB2005, TimmPRB2008, JinJCP2008, KollerPRB2010}.

We consider here approximate master equation approaches, which were also used to interpret transport experimental data in the regime of weak coupling to the environment \cite{NilssonPRL2010,HuttelPRL2009,ZyazinNL2010}. In particular, we address the so-called \textit{Pauli} master equation \cite{BreuerBook2006, PauliBook1928, BruusBook2004}, the \textit{first-order Redfield} approach \cite{BreuerBook2006,WangsnessPR1953,RedfieldIBM1957}, the \textit{first/second-order von Neumann} approaches \cite{PedersenPRB2005,PedersenPHE2010}, and a particular form of the \textit{Lindblad} equation \cite{LindbladCMP1976,Kirsanskas2017Lindblad}, and apply these methods for tunneling models, where the quantum dots can have arbitrary Coulomb interactions \cite{MeirPRL1993}. Even though there is a lot of literature on such methods, there is no publicly-available, well-documented package, which is easy to apply for quantum dot model systems. One reason is that depending on parameter regime these methods can fail and a good knowledge of the method is required to know whether to trust the result or not. For example, these methods can violate positivity \cite{BreuerBook2006} of the reduced density matrix and lead to large currents flowing against the bias \cite{GoldozianSciRep2016}. Nevertheless, we think it is important to have a package where a user can duplicate existing calculations, check the applicability of different methods, or simply discover new kind of physics using different approximate master equations.

In this paper, we describe an open-source package \texttt{QmeQ} designed for simulating stationary-state transport through quantum dots (available at \url{http://github.com/gedaskir/qmeq}). The name \texttt{QmeQ} deciphers as \textit{\textbf{Q}uantum} \textit{\textbf{m}aster} \textit{\textbf{e}quation} for \textit{\textbf{Q}uantum} dot transport calculations. It is written in the programming languages \texttt{Python} \cite{Python} and \texttt{Cython} \cite{Cython,BehnelComputSciEng2011}. In recent years, a large number of \texttt{Python}-based scientific computation packages have appeared \cite{SciPyTopical} because of the development of \texttt{NumPy} \cite{VanderWaltComputSciEng2011} and \texttt{SciPy} \cite{SciPy}. These have modules for linear algebra, optimisation, and special functions, which also \texttt{QmeQ} relies on. Thus it is convenient to use the same open-source platform based on \texttt{Python}, which provides an easy integration between different kind of packages. Additionally, \texttt{Python} is a high-level interpreted language, which allows development of an easy-to-read code and easy usage. For some parts of the code we use \texttt{Cython} which lets us to compile bottleneck parts of the package to achieve better performance.\footnote{In short, \texttt{Cython} converts \texttt{Python}-like code with static type definitions into \texttt{C} code.} Furthermore, in our examples for data visualization we use the \texttt{matplotlib} package \cite{HunterComputSciEng2007}, the tutorials are provided using \texttt{Jupyter} notebooks \cite{PerezComputSciEng2007}, and detailed documentation of the code is generated using \texttt{Sphinx} \cite{Sphinx}.

There are already a variety of well-developed open-source \texttt{Python} packages dedicated to model quantum mechanical phenomena and transport. One such package is \texttt{ASE} (Atomic Simulation Environment), which performs atomistic simulations with the possibility to make transport calculations in molecular junctions using density-functional theory (DFT) and non-equilibrium Green's functions (NEGF) \cite{BahnComputSciEng2002}. For numerical calculations on tight-binding models with a focus on quantum transport the \texttt{Kwant} package can be used \cite{GrothNJP2014}. The simulation of the dynamics of open quantum systems using master equations in Lindblad, Floquet–Markov, or Bloch–Redfield form can be performed using the \texttt{QuTiP} (Quantum Toolbox in Python) framework \cite{JohanssonComputPhysCommun2012,JohanssonComputPhysCommun2013}.
In contrast to \texttt{QuTiP}, we focus on a particular model for the quantum dot device, where Coulomb interaction between particles can be dominating, and allow for a direct comparison between different approximate quantum master equations, including the \emph{second-order von Neumann} approach.
%
%The difference between \texttt{QuTiP} and our package \texttt{QmeQ} is that we focus on . Then for the particular microscopic Hamiltonian different approximate quantum master equation approaches can be applied
%non-equilibrium transport through quantum system, which are coupled to the environment by tunnel Hamiltonian where the Coulomb interaction between particles is dominating.
%
For an overview of other transport packages including commercial ones also see Section~4 of Ref.~\cite{GrothNJP2014}. Lastly, the cotunneling calculations using a $T$-matrix based approach \cite{BruusBook2004} can be performed using the \texttt{humo} package.\footnote{Available at \url{http://github.com/georglind/humo}. This package is currently rather undocumented but the used methods are well described in Ref.~\cite{PedersenPhd2013}.  The calculations can be performed on models of the type described by Eq.~\eqref{ham}. }

The paper is organized as follows. In Section~\ref{sec:model} we introduce the model Hamiltonian for the quantum dot device and in Section~\ref{sec:qmeqmodel} we describe how such a Hamiltonian is set up in \texttt{QmeQ}. In Sections~\ref{sec:ssorb}, \ref{sec:sdqdot}, and \ref{sec:sstd} we give examples of transport calculation through various quantum dot devices, using different approaches (examples scripts are available at \url{http://github.com/gedaskir/qmeq-examples}). The derivation of the approximate master equations used in \texttt{QmeQ} is relegated to Appendices. %~\ref{App}
Throughout the paper our units are such that $\hbar=1$, $k_{\mr{B}}=1$, $\abs{e}=1$.

\section{\label{sec:model}The model}

The system under consideration consists of a number of quantum dots attached to a number of metallic leads. Such a quantum dot device can be modelled by the following Hamiltonian:
\begin{equation}\label{ham}
H=H_{\mr{leads}}+H_{\mr{tunneling}}+H_{\mr{dot}},
\end{equation}
where $H_{\mr{leads}}$ describes electrons in the leads, $H_{\mr{dot}}$ describes electrons in the dot, and $H_{\mr{tunneling}}$ corresponds to tunneling between the leads and the dot. The leads are described as non-interacting electrons
\begin{equation}\label{hamLeads}
H_{\mr{leads}}=\sum_{\leadqn k}\leadE{\leadqn k}^{\phantom{\dagger}}\cd_{\leadqn k}\can_{\leadqn k},
\end{equation}
where $\cd_{\leadqn k}$ creates an electron in the lead channel $\leadqn$ with $k$ representing a continuous quantum number, which refers to a continuum energy. This means that $k$-sums can be performed by introducing the density of states $\nu(E)$ as $\sum_{k}f_{k}\rightarrow\int \mr{d}E\nu(E)f(E)$. The lead channel $\leadqn$ can encompass, for example, an actual source/drain lead label, spin of an electron, etc., depending on actual physical setup under consideration.

For the quantum dot the following rather general many-body Hamiltonian is used
%
%\begin{subequations}
\begin{equation}\label{hamQD}
H_{\mr{dot}}=H_{\mr{single}}+H_{\mr{Coulomb}},
\end{equation}
\begin{equation}\label{hamQDsingle}
H_{\mr{single}}=\sum_{i}\varepsilon_{i}^{\phantom{\dagger}}\dd_{i}\dan_{i}
+\sum_{i\neq j}\Omega_{ij}^{\phantom{\dagger}}\dd_{i}\dan_{j},
\end{equation}
\begin{equation}\label{hamQDcoulomb}
H_{\mr{Coulomb}}=\sum_{mnkl}U_{mnkl}\dd_{m}\dd_{n}\dan_{k}\dan_{l},
\quad\text{with } m<n,
\end{equation}
%\end{subequations}
%
where $\dd_{i}$ creates an electron in single-particle orbital $i$, $\varepsilon_{i}$ is the energy of that orbital, and $\Omega_{ij}$ gives the hybridization between single-particle orbitals. Here also the Coulomb interaction $U_{mnkl}$ can be present. We note that the Coulomb interaction takes the form of Eq.~\eqref{hamQDcoulomb} (without a factor of $\tfrac{1}{2}$) for state labels $m<n$. Lastly, the tunneling Hamiltonian is
\begin{equation}\label{hamT}
H_{\mr{tunneling}}=\sum_{\leadqn k i}t_{\leadqn k, i}^{\phantom{\dagger}}\dd_{i}\can_{\leadqn k}+\mr{H.c.},
\end{equation}
where $\mr{H.c.}$ denotes Hermitian conjugate of the first term and $t_{\leadqn k, i}$ is the tunneling amplitude between the leads and the dot. An important energy scale in the calculations is the tunneling rate defined as
\begin{equation}
\Gamma_{\leadqn k, i}(E)=2\pi\sum_{k}\lvert{t_{\leadqn k,i}}\rvert^2\delta(E-\varepsilon_{\leadqn k}).
\end{equation}
In the case of single spinful orbital with on-site interaction $U$ Hamiltonian Eq.~\eqref{ham} is referred to as Anderson-type model \cite{MeirPRL1993,AndersonPR1961,HaugBook1998}.

\subsection{Approximations}

For the calculations in \texttt{QmeQ} we make the following important assumption for the model Eq.~\eqref{ham}:
\begin{enumerate}
\item[\textbf{1.}] The leads are thermalized according to a Fermi-Dirac occupation function $f_{\leadqn}(E)=[e^{(E-\mu_{\leadqn})/T_\leadqn}+1]^{-1}$ with respective temperatures $T_{\leadqn}$ and chemical potentials $\mu_{\leadqn}$.
\end{enumerate}
Additionally, we use the so-called wide-band limit for the leads:
\begin{enumerate}
\item[\textbf{2.}] The leads have constant density of states $\nu(E)\approx\nu(E_{F})=\nu_{F}$, where the subscript $F$ corresponds to the Fermi level. Then the $k$-sums are performed as $\sum_{k}\rightarrow\nu_{F}\int_{-D}^{+D}\mr{d}E$, where $D$ denotes bandwidth of the leads. Also the tunneling amplitudes are energy (or $k$) independent, i.e., $t_{\leadqn k, i}\approx t_{\leadqn i}$. In this case the tunneling rates become $\Gamma_{\leadqn i}=2\pi\nu_{F}\lvert{t_{\leadqn i}}\rvert^2$.
\end{enumerate}
Approximation \textbf{2.} is made for convenience to simplify the integrals and not to clutter the \texttt{QmeQ} code with the specification of energy-dependent functions. The wide-band limit can be easily relaxed. On the other hand, this is often a good %a valid
approximation for metallic leads.

\subsection{Many-body eigenstates and the reduced density matrix}
In \texttt{QmeQ} the quantum dot Hamiltonian \eqref{hamQD} is constructed in a Fock basis and diagonalized exactly to obtain the many-body eigenstates $\ket{a}$,
\begin{equation}\label{hamQDdiag}
H_{\mr{dot}}=\sum_{a}E_{a}\ket{a}\bra{a}.
\end{equation}
We note that a fermionic many-body Hamiltonian is constructed efficiently using Lin tables (for more details see Refs.~\cite{PedersenPhd2013,LinComputPhys1993}).
%(see Section~\ref{sec:siDM} for more details)
%
The tunneling Hamiltonian is expressed in this many-body eigenbasis as
\begin{equation}\label{hamT2}
\begin{aligned}
&H_{\mr{tunneling}}=\sum_{ab,\leadqn k}T_{ba,\leadqn}\ket{b}\bra{a}c_{\leadqn k}+\mr{H.c.},\quad
T_{ba,\leadqn}=\sum_{i}t_{\leadqn i}\bra{b}\dd_{i}\ket{a},
\end{aligned}
\end{equation}
where $T_{ba,\leadqn}$ represents many-particle tunneling amplitudes. Here we used the \textit{letter convention}: if more than one state enters an equation, then the position of the letter in the alphabet follows the particle number (for example $N_b= N_a +1$, $N_c= N_a +2$, $N_{a'}=N_a$). In such a way the sum $\sum_{bc}$ restricts to those combinations, where $N_c = N_b+1$. The approximate quantum master equations are constructed in the many-body eigenbasis of the quantum dot and are discussed in Appendices.

The central quantity in the approximate master equations is the reduced density matrix, which is defined as: % [also see Eq.~\eqref{phieqs}]
\begin{equation}
\Phi_{bb'}^{[0]}=\sum_{g}\bra{bg}\rho\ket{b'g},
\end{equation}
%
%with $\rho$ being the full density matrix operator of the system. Here $g$ represents many-body eigenstate of the lead Hamiltonian~$H_{\mr{leads}}$~\eqref{hamLeads} with $\ket{bg}=\ket{b}\otimes\ket{g}$.
with $\rho$ being the full density matrix operator of the system. Here $g$ represents the many-body eigenstates of the lead Hamiltonian~$H_{\mr{leads}}$~\eqref{hamLeads} with arbitrary occupations of the lead states $k\alpha$. Then a basis of the combined lead and dot system is given by the product states $\ket{bg}=\ket{b}\otimes\ket{g}$.

\subsection{Calculation of currents}

In \texttt{QmeQ} the current emanating from a particular lead channel $\leadqn$ is calculated for a stationary state. The particle current leaving the lead channel $\leadqn$ is defined as
\begin{equation}\label{general_currentDef}
I_{\leadqn}= -\frac{\partial}{\partial t}\avgs{N_{\leadqn}} = -\ii\avgs{[H, N_{\leadqn}]},
\quad \text{with} \quad
N_{\leadqn}=\sum_{k}\cd_{\leadqn k}\can_{\leadqn k},
\end{equation}
and the energy current as
\begin{equation}\label{general_heatcurrentDef}
\dot{E}_{\leadqn} = -\frac{\partial}{\partial t}\avgs{H_{\leadqn}} = -\ii\avgs{[ H, H_{\leadqn}]},
\quad \text{with} \quad
H_{\leadqn}=\sum_{k}\leadE{\leadqn k}^{\phantom{\dag}}\cd_{\leadqn k}\can_{\leadqn k}.
\end{equation}
%
%The $\Phi^{[1]}$'s introduced in Eq.~\eqref{phieqs} represent
We introduce the following energy-resolved particle-current amplitudes:
\begin{equation}
\Phi_{cb,\alpha k}^{[1]}=\sum_{g}
\bra{cg-\alpha k}\rho\ket{bg}(-1)^{N_{b}},
\quad\text{with}\quad
\ket{bg-\alpha k}=\ket{b}\otimes\can_{\alpha k}\ket{g},
\end{equation}
where $N_{b}$ is the number of particles in the many-body state $\ket{b}$.
%
%Using  $\Phi^{[1]}$'s
We can express the particle and energy currents as
\begin{align}\label{cureq}
&I_{\leadqn}=-2\sum_{k,cb}\Imag[T_{bc,\leadqn}^{\phantom{[1]}}\Phi_{cb,\leadqn k}^{[1]}],\\
&\dot{E}_{\leadqn}=-2\sum_{k,cb}\leadE{\leadqn k}\Imag[T_{bc,\leadqn}\Phi_{cb,\leadqn k}^{[1]}].
\end{align}
Lastly, the heat current emanating from the lead channel $\leadqn$ can be defined as
\begin{equation}
\dot{Q}_{\leadqn}=\dot{E}_{\leadqn}-\mu_{\leadqn}I_{\leadqn}.
\end{equation}
The above definition of the heat current corresponds to the rate of change of the entropy $\dot{S}_{\leadqn}=\tfrac{1}{T_{\leadqn}}(\dot{E}_{\leadqn}-\mu_{\leadqn}\dot{N}_{\leadqn})$.

\section{\label{sec:qmeqmodel}The \texttt{QmeQ} package}

\begin{figure}[!ht]
\begin{center}
\includegraphics[width=0.60\textwidth]{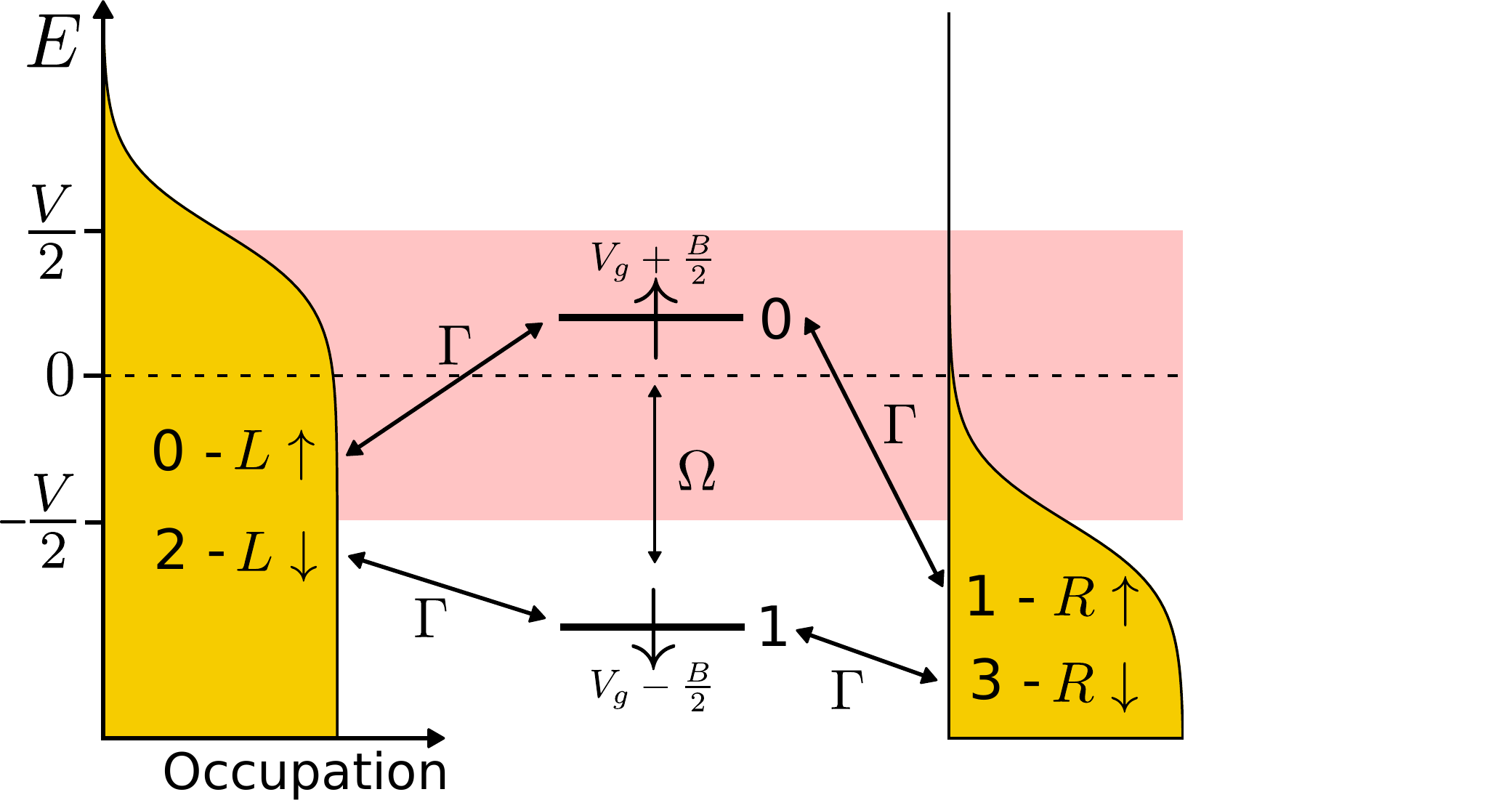}
\caption{Spinful single orbital coupled to spinful metallic leads, where $\Gamma_{L}$ and $\Gamma_{R}$ are the tunneling rates to the source ($L$) and the drain ($R$) leads, respectively. Numbers represent the labeling of states in the example. Energies of the spin-up and spin-down levels are given by $\ve_{0}$ and $\ve_{1}$, respectively. $\Omega$ is the hybridisation between the states, which can appear due to presence of spin-orbit coupling (sets a concrete spin quantization axis) and a magnetic field perpendicular to the spin orientation. }
\label{sorb}
\end{center}
\end{figure}

The \texttt{QmeQ} package diagonalises the quantum dot Hamiltonian \eqref{hamQD} exactly and performs various approximate master equation calculations on the system Hamiltonian \eqref{ham}, where tunneling \eqref{hamT} is treated as a perturbation. Installation instructions for \texttt{QmeQ} are provided in the \href{https://github.com/gedaskir/qmeq/tree/master/INSTALL.md}{INSTALL.md} file coming with the source code of the package. We start with a short script giving a minimal example of a quantum dot containing one spinful orbital and on-site charging energy $U$ coupled to source ($L$) and drain ($R$) leads as shown in \figurename~\ref{sorb}:
\begin{equation}\label{hamOne}
H_{\mr{one}}=\sum_{\substack{k;\ell=L,R \\ \sigma=\up,\down}}\ve_{\ell k}^{\phantom{\dagger}}\cd_{\ell\sigma k}\can_{\ell\sigma k}
+\sum_{\ell\sigma k}\left(t_{\ell}\dd_{\sigma}\can_{\ell\sigma k}+\mr{H.c.}\right)
+\sum_{\sigma}\ve_{\sigma}\dd_{\sigma}\dan_{\sigma}
+(\Omega \dd_{\up}\dan_{\down}+\mr{H.c.})
+U\dd_{\up}\dd_{\down}\dan_{\down}\dan_{\up}.
\end{equation}
The script calculating the particle current and the energy current in a single point in the parameter space reads
\begin{python}
# Prerequisites
import qmeq
from numpy import sqrt, pi, power

# Parameters used (numbers are in arbitrary energy_scale, e.g., meV)
e0, e1, omega, U = 0.0, 0.0, 0.0, 20.0
# dband is bandwidth D of the leads
tempL, tempR, muL, muR, dband= 1.0, 1.0 , 0.2, -0.2, 60.0
gammaL, gammaR = 0.5 , 0.7
tL, tR = sqrt(gammaL/(2*pi)), sqrt(gammaR/(2*pi))

## Hamiltonian, which requires definition

# nsingle is the number of single-particle states
# Here level 0 is spin up; level 1 is spin down
nsingle = 2
hsingle = {(0,0): e0, (1,1): e1, (0,1): omega}
# (0,1,1,0) represents an operator d0^{+}d1^{+}d1^{-}d0^{-}
coulomb = {(0,1,1,0): U}

## Lead and tunneling properties, which requires definition
# nleads is the number of lead channels
nleads = 4 # for two leads L/R with two spins up/down
# Here   L-up      R-up      L-down    R-down
mulst = {0: muL,   1: muR,   2: muL,   3: muR}
tlst =  {0: tempL, 1: tempR, 2: tempL, 3: tempR}

# The coupling matrix has indices(lead-spin, level)
tleads = {(0,0): tL, (1,0): tR, (2,1): tL, (3,1): tR}

## Construction of the transport system and
## calculation of currents into the system from all 4 channels

# Choice of approximate approach
# For kerntype='Redfield', '1vN', 'Lindblad', or 'Pauli'
system = qmeq.Builder(nsingle, hsingle, coulomb,
                      nleads, tleads, mulst, tlst, dband,
                      kerntype='Pauli')
system.solve()
print('Output is currents in four lead channels')
print('[L-up R-up L-down R-down]\n')
print('Pauli, particle current (in units of energy_scale/hbar):')
print(system.current)
print('Pauli, energy current (in units of energy_scale^2/hbar):')
print(system.energy_current)

# For '2vN' approach
# Here we use 2^12 energy grid points for the leads
kpnt = power(2,12)
system2vN = qmeq.Builder(nsingle, hsingle, coulomb,
                         nleads, tleads, mulst, tlst, dband,
                         kerntype='2vN', kpnt=kpnt)
# 7 iterations is good for temp approx gamma
system2vN.solve(niter=7)
print('\n2vN, particle current (in units of energy_scale/hbar):')
print(system2vN.current)
print('2vN, energy current (in units of energy_scale^2/hbar):')
print(system2vN.energy_current)
\end{python}
Here we have calculated the particle current and the energy current from the four lead channels $L\uparrow$, $R\uparrow$, $L\downarrow$, $R\downarrow$ into the dot using \emph{Pauli} master equation and \emph{second-order von Neumann} approach (\emph{2vN}) and we get the following output:
\begin{outsty}
Output is currents in four lead channels
[L-up R-up L-down R-down]

Pauli, particle current (in units of energy_scale/hbar):
[ 0.01948779 -0.01948779  0.01948779 -0.01948779]
Pauli, energy current (in units of energy_scale^2/hbar):
[  1.59585694e-09  -1.59585694e-09   1.59585694e-09  -1.59585694e-09]

2vN, particle current (in units of energy_scale/hbar):
[ 0.01431084 -0.01431084  0.01431084 -0.01431084]
2vN, energy current (in units of energy_scale^2/hbar):
[ 0.00617130 -0.00617129  0.00617130 -0.00617129]
\end{outsty}
We elaborate more on this single spinful orbital example in Section~\ref{sec:ssorb}. Next we discuss what different parts of the script mean.

From the above example we see that the system is defined using the \pyin{Builder} class. For \pyin{Builder} to construct a system object, we need to specify at least the following variables:
\begin{pythonpseudo}
system = qmeq.Builder(nsingle, # Number of single-particle states
                      hsingle, # Single-particle Hamiltonian
                      coulomb, # Coulomb matrix elements
                      nleads,  # Number of lead channels
                      tleads,  # Single-particle tunneling amplitudes
                      mulst,   # Chemical potentials of leads
                      tlst,    # Temperatures of leads
                      dband)   # Bandwidth of leads
\end{pythonpseudo}
Additionally, there is a variety of optional arguments, which can be viewed by typing \pyin{help(qmeq.Builder)} in the \texttt{Python} interpreter. These optional arguments can be set by specifying {\ttm Builder(..., }\emph{argument=value}{\ttm)}.\footnote{For the \emph{2vN} approach it is necessary to set {\ttmfoot kpnt}, which is the number of the equidistant energy grid points for the lead electrons.}

The variables \pyin{nsingle}, \pyin{hsingle}, and \pyin{coulomb} describe the quantum dot Eq.~\eqref{hamQD}. \pyin{nsingle} is the number of single-particle states. Single-particle states are labeled by integers from $0$ to \pyin{nsingle}$-1$, and the single-particle Hamiltonian \eqref{hamQDsingle} \pyin{hsingle} is specified using the \texttt{Python} dictionary:
%The default specification is using \texttt{Python} dictionary and it has the following format:
%
\begin{pythonpseudo}
hsingle = {($0$,$0$): $\ve_0$, ($0$,$1$): $\Omega_{01}$, $\ldots$ , ($0$,nsingle$-1$): $\Omega_{0,\text{\ttmsub nsingle}-1}$,
           $\ldots$
           ($i$,$i$): $\ve_{i}$, ($i$,$i+1$): $\Omega_{12}$, $\ldots$ , ($i$,nsingle$-1$): $\Omega_{i,\text{\ttmsub nsingle}-1}$,
           $\ldots$
           (nsingle-1,nsingle-1): $\Omega_{\text{\ttmsub nsingle}-1,\text{\ttmsub nsingle}-1}$}
\end{pythonpseudo}
In a dictionary it is enough to specify one element like $\Omega_{ij}\dd_{i}\dan_{j}$, because the element $\Omega_{ji}\dd_{j}\dan_{i}$ is determined by complex conjugation $\Omega_{ji}=\Omega_{ij}^{*}$ and is included automatically in order to get a Hermitian quantum dot Hamiltonian. If an element like \pyin{($j$, $i$)} is given, it will be added to the Hamiltonian. So specifying %{\ttm(}$i${\ttm,}$j${\ttm):}
\pyin{ \{($i$,$j$):$\Omega_{ij}$, ($j$,$i$):$\Omega_{ij}^{*}$\} } will simply double count $\Omega_{ij}$. Default values of $\ve_{i}$ and $\Omega_{ij}$ are zero.

The interaction Hamiltonian \eqref{hamQDcoulomb} \pyin{coulomb} can be specified using a dictionary as
\begin{pythonpseudo}
coulomb = {($m$,$n$,$k$,$l$): $U_{mnkl}$,
           $\ldots$
           ($i$,$j$,$p$,$r$): $U_{ijpr}$}
\end{pythonpseudo}
where $U_{mnkl}\dd_{m}\dd_{n}\dan_{k}\dan_{l}$ and $U_{ijpr}\dd_{i}\dd_{j}\dan_{p}\dan_{r}$ represent some non-zero Coulomb interaction matrix elements.

The variable \pyin{nleads} is an integer and sets the number of lead channels (we will simply call it leads). The properties of leads can be given in dictionaries \pyin{mulst} (chemical potential) and \pyin{tlst} (temperature), which have \pyin{nleads} entries. Also the leads are labeled by integers from $0$ to \pyin{nleads}$-1$. So the lead properties can be set by using
\begin{pythonpseudo}
mulst = {$0$: $\mu_{0}$, $1$: $\mu_{1}$, $\ldots$, nleads$-1$: $\mu_{\text{\ttmsub nleads}-1}$}
tlst  = {$0$: $T_{0}$, $1$: $T_{1}$, $\ldots$, nleads$-1$: $T_{\text{\ttmsub nleads}-1}$}
dband = $D$
\end{pythonpseudo}
We note that \pyin{dband} is a floating point number representing the bandwidth $D$ of the leads.

Lastly, the single-particle tunneling amplitudes \pyin{tleads} can be specified using a dictionary as
\begin{pythonpseudo}
tleads = {($\leadqn$,$i$): t$_{\leadqn i}$,
          $\ldots$
          ($\beta$,$j$): t$_{\beta j}$}
\end{pythonpseudo}
where first index in ($\leadqn$, $i$) is the lead label and second index is the single-particle state label. Here $\text{\ttm t}_{\leadqn i}=\sqrt{\nu_{F}}t_{\leadqn i}$.

\phantom{...}

\phantom{...}

Once the \pyin{system} object was constructed using the \pyin{Builder} class it can be solved by simply using \footnote{For the \emph{2vN} approach it is also mandatory to specify the number of iterations, {\ttmfoot system.solve(niter={\footnotesize\emph{integer}})}.}
\begin{pythonpseudo}
system.solve()
\end{pythonpseudo}
The \pyin{system.solve()} performs the following steps
\begin{enumerate}
\item[\textbf{1.}] \pyin{qdq}, obtains many-body eigenstates and eigenenergies of the quantum dot Hamiltonian \eqref{hamQD} (sets \pyin{system.qd.Ea}).
\item[\textbf{2.}] \pyin{rotateq}, expresses the tunnelling Hamiltonian \eqref{hamT} in many-body eigenbasis (sets \pyin{system.lead.Tba}).
\item[\textbf{3.}] \pyin{masterq}, solves approximate master equation and obtains reduced density matrix elements $\Phi^{[0]}$ (sets \pyin{system.phi0}).
\item[\textbf{4.}] \pyin{currentq}, calculates the current amplitudes and currents (sets \pyin{system.phi1}, \pyin{system.current}, etc.).
\end{enumerate}
Unnecessary steps can be skipped. For example,
\begin{pythonpseudo}
system.solve(qdq=True, rotateq=True, masterq=False, currentq=False)
\end{pythonpseudo}
will perform step \textbf{1.} (\pyin{qdq=True}) and \textbf{2.} (\pyin{rotateq=True}), however, the master equation will not be solved (\pyin{masterq=False}) and the current will not be calculated (\pyin{currentq=False}).\footnote{For the \emph{2vN} approach, if {\ttmfoot masterq=True} then the current is always calculated.}

How step \textbf{3.} is performed depends on the master equation used. For first-order approaches we solve the following linear equations
\begin{align}
\label{liouveq}\mc{L}\Phi^{[0]}&=0,\\
\label{liouveqnorm}\Tr[\Phi^{[0]}]&=1,
\end{align}
where $\mc{L}$ is the \textit{kernel} (or \textit{Liouvillian}) of the approach (see Appendices for more details). The solution procedure for the \emph{2vN} approach is more complicated and is described in \ref{App:2vN}.

After the calculation different properties of the system can be accessed through
\begin{pythonpseudo}
system.current        # Particle current
system.energy_current # Energy current
system.heat_current   # Heat current
system.phi0           # Reduced density matrix elements
system.phi1           # Energy integrated current amplitudes
system.Ea             # Eigenenergies of the quantum dot
system.Tba            # Many-body tunneling amplitudes
\end{pythonpseudo}
For example, \pyin{system.current} is a one-dimensional \texttt{NumPy} array of length \pyin{nleads}.

\subsection{Attributes and functions of the {\ttm Builder} class}
Here we concisely describe some of the useful attributes and functions of the {\ttm Builder} class. The variable
\begin{pythonpseudo}
system.indexing # Possible values: 'Lin', 'charge', 'sz', 'ssq'
\end{pythonpseudo}
determines the labeling of states and symmetries, which can be used to simplify the numerical calculations. We note that \pyin{'sz'} stands for classification using total spin projection $S_{z}$ and \pyin{'ssq'} stands for $S^2$. For \pyin{'Lin'} and \pyin{'charge'} only the charge is considered as a good quantum number. For now the usage of symmetries to optimize the calculations is available only for the first-order approaches in \texttt{QmeQ}.
The central idea of labeling of many-body states is so-called Lin tables \cite{PedersenPhd2013,LinComputPhys1993}. Here Fock states are labeled by a decimal number and an actual Fock state is obtained by converting the decimal label to a binary number. For example, let us say we have \pyin{nsingle=6}, which leads to \pyin{nmany=64}. Then the labels `$33$' and `$44$' would represent different Fock states for different \pyin{indexing}
%33 [1, 0, 0, 0, 0, 1] [1, 0, 0, 1, 0, 1] [0, 1, 1, 0, 1, 0]
%44 [1, 0, 1, 1, 0, 0] [0, 1, 1, 0, 1, 1] [1, 0, 0, 1, 1, 1]
%
\begin{pythonpseudo}
      'Lin'      'charge'    'sz'$\text{ or }$'ssq'
 $33:$ $\ket{1,0,0,0,0,1},\quad$ $\ket{1,0,0,1,0,1},\quad$ $\ket{0_{\up},1_{\up},1_{\up},0_{\down},1_{\down},0_{\down}}$
 $44:$ $\ket{1,0,1,1,0,0},\quad$ $\ket{0,1,1,0,1,1},\quad$ $\ket{1_{\up},0_{\up},0_{\up},1_{\down},1_{\down},1_{\down}}$
\end{pythonpseudo}
Decimal numbers $33$ and $44$ are the binary numbers $100001$ and $101100$. For \pyin{'charge'} indexing the Lin labels are sorted by increasing charge of the states and for \pyin{'sz'} and \pyin{'ssq'} the sorting goes first by the charge and then by the $S_{z}$ value of states.

The quantum dot Hamiltonian has \pyin{nmany=$2^{\text{\ttmsub nsingle}}$} many-body eigenstates $\ket{b}$, which in Eq.~\eqref{hamQDdiag} are labeled by integers $b$ from $0$ to \pyin{nmany}$-1$. To get information about the many-body eigenstate $\ket{b}$ use the function
\begin{pythonpseudo}
system.print_state(b)
\end{pythonpseudo}
and to get information about the all many-body eigenstates use
\begin{pythonpseudo}
system.print_all_states(filename)
\end{pythonpseudo}
where \pyin{filename} is a string giving the file destination where this information will be printed. After diagonalisation depending on \pyin{indexing}, the many-body eigenstates have a particular labeling convention. For example, with $\pyin{indexing='ssq'}$, which has the highest symmetry, the states are sorted in the following order of parameters: charge, $S_{z}$, $S^2$, and energy $E_{b}$. To change the sorting order and relabel the states use the function
\begin{pythonpseudo}
system.sort_eigenstates(srt=[$l_0$,$l_1$,$l_2$,$l_3$])
\end{pythonpseudo}
Here $l_{i}\in\{0,1,2,3\}$, where different numbers represent such properties: $0$ -- $E_{b}$, $1$ -- charge, $2$ -- $S_{z}$, $3$ -- $S^{2}$. So the default sorting for \pyin{'ssq'} is \pyin{srt=[1,2,3,0]}. For example, in order to sort by the energy and then by the charge use \pyin{srt=[0,1]}. However, we note that \pyin{sort_eigenstates} affects the labels when printing the states and the default labels are used in actual calculations. For instance, the order of energies in \pyin{system.Ea} or tunneling amplitudes \pyin{system.Tba} will not be changed by using \pyin{sort_eigenstates}.

In some master equation calculations not all many-body states may be relevant and contribute to the transport, because they are very far above the ground state in energy. To neglect the states with energy higher than $\Delta{E}$ above the ground state use
\begin{pythonpseudo}
system.remove_states($\Delta{E}$)
\end{pythonpseudo}
and to reset the usage of all states call
\begin{pythonpseudo}
system.use_all_states()
\end{pythonpseudo}

To access particular matrix elements of the reduced density matrix $\Phi_{bb'}^{[0]}$ and current amplitudes $\Phi_{cb,\leadqn}^{[1]}$ the following function can be used
\begin{pythonpseudo}
system.get_phi0($b$, $b'$)
system.get_phi1($\leadqn$, $b$, $b'$)
\end{pythonpseudo}
Values of $\Phi_{bb'}^{[0]}$ and $\Phi_{cb,\leadqn}^{[1]}$ are stored in the arrays
\begin{pythonpseudo}
system.phi0
system.phi1
\end{pythonpseudo}
We note that there is a difference how the reduced density matrix $\Phi^{[0]}$ is stored in \pyin{phi0} for the first-order approaches and the \emph{2vN} approach. For the \emph{2vN} approach $\pyin{phi0}$ is a one-dimensional array of \pyin{complex} numbers of length \pyin{ndm0}, where \pyin{ndm0} denotes the total number of matrix elements in $\Phi^{[0]}$. For the first-order approaches we exploit the Hermiticity $\Phi_{b'b}^{[0]}=\Phi_{bb'}^{[0]*}$ and store just the elements with $b\leq b'$, where $\pyin{ndm0}$ now denotes the number of complex elements in the upper triangle of $\Phi^{[0]}$. In this case \pyin{phi0} is a one-dimensional array of \pyin{float} numbers and is of length \pyin{2*ndm0-npauli}, where \pyin{npauli} is the number of diagonal elements. The diagonal elements $\Phi_{bb}^{[0]}$ can be accessed by \pyin{phi0[0:npauli]}. The entries \pyin{phi0[npauli:ndm0]} and \pyin{phi0[ndm0:]} contain the real and imaginary parts of $\Phi_{bb'}^{[0]}$ with $b<b'$. The variables \pyin{npauli} and \pyin{ndm0} are contained in
\begin{pythonpseudo}
system.si.npauli
system.si.ndm0
\end{pythonpseudo}
The Liouvillian $\mc{L}$ corresponding to the reduced density matrix $\Phi^{[0]}$ can be accessed through
\begin{pythonpseudo}
system.kern
\end{pythonpseudo}

Lastly, when doing calculations there is a need to change the parameters of the system. This can be achieved by using the functions
\begin{pythonpseudo}
system.add(hsingle, coulomb, tleads, mulst, tlst)
system.change(hsingle, coulomb, tleads, mulst, tlst)
\end{pythonpseudo}
where as the names suggest \pyin{system.add} and \pyin{system.change} adds and changes the value of a parameter. For example, to change the temperature of the lead $0$ to $T_{0}$ one can use the call: \pyin{system.change(tlst=\{0: $T_{0}$\})}. To modify \pyin{dband} to value $D$ simply use \pyin{system.dband=$D$}.

\subsection{Important classes}

\begin{figure}[t]
\begin{center}
\includegraphics[width=0.80\textwidth]{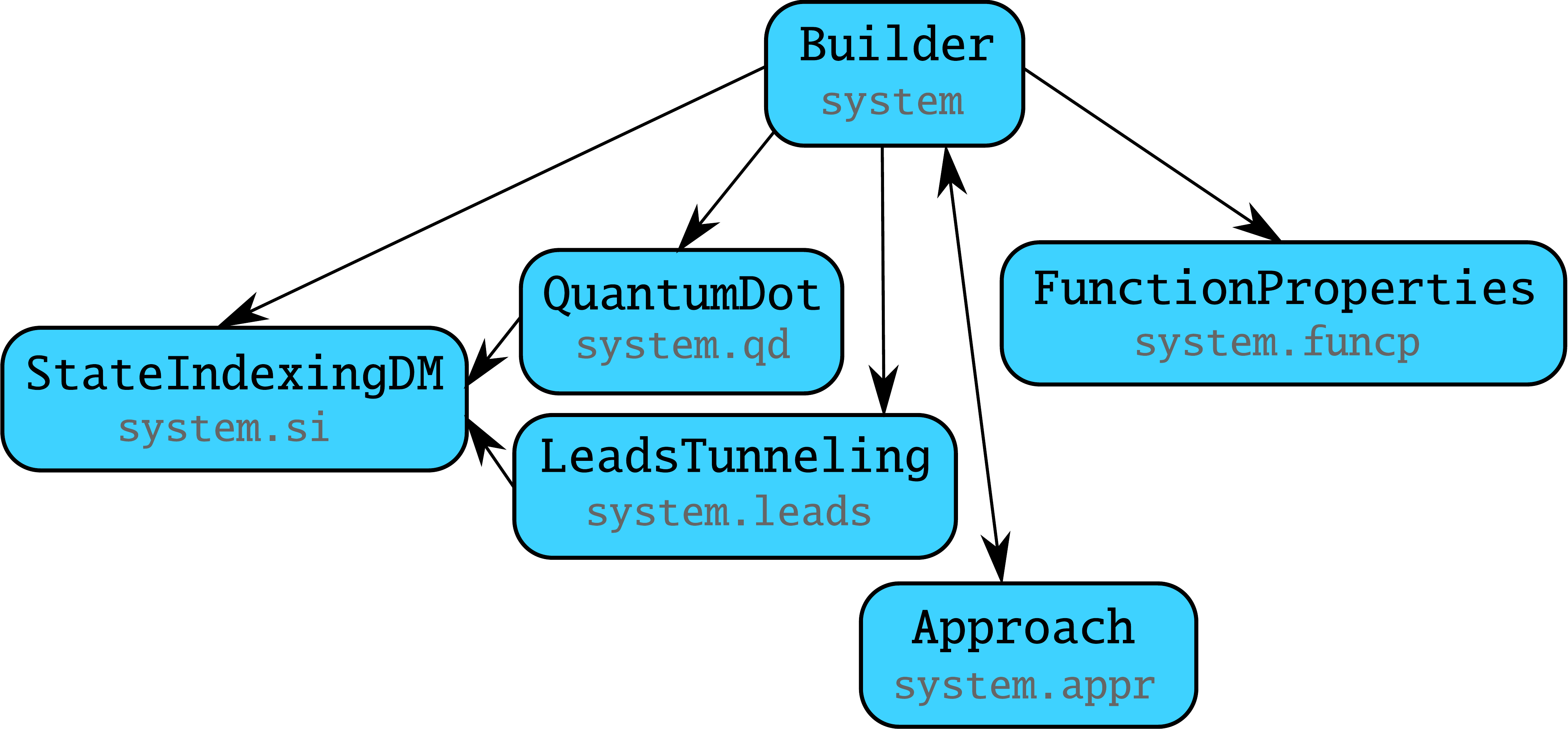}
\caption{Important classes and their interrelation in \texttt{QmeQ} package. The direction of arrow $A\rightarrow B$ means that class $A$ is dependent on class $B$.}
\label{impcls}
\end{center}
\end{figure}

\texttt{QmeQ} \pyin{Builder} interacts with different classes to set up the system and perform calculations using different approaches as shown in \figurename~\ref{impcls}. Here we concisely describe these classes. \pyin{StateIndexingDM} is a class describing the indexing of many-body states and density matrix elements. An object of this class is accessed in previously constructed \pyin{Builder} object {\ttm system} through
\begin{pythonpseudo}
system.si
\end{pythonpseudo}
The class \pyin{StateIndexingDMc} is used for the \emph{2vN} approach. Both \pyin{StateIndexingDM} and  \pyin{StateIndexingDMc} are derived from \pyin{StateIndexing}. The \pyin{QuantumDot} is a class used internally in the \pyin{Builder} to construct and diagonalise the many-body Hamiltonian \eqref{hamQD}. An object of this class is accessed through
\begin{pythonpseudo}
system.qd
\end{pythonpseudo}
The \pyin{LeadsTunneling} is a class used to represent properties of the leads, construct many-body tunneling amplitudes, and express it the eigenbasis of the Hamiltonian. An object of this class is accessed through
\begin{pythonpseudo}
system.leads
\end{pythonpseudo}
The \pyin{FunctionProperties} is a class containing miscellaneous variables used by the \pyin{Approach} class. An object of this class is accessed through
\begin{pythonpseudo}
system.funcp
\end{pythonpseudo}
Finally, the classes which deal with implementation of different approaches are derived from \pyin{Approach} class, and its object is
\begin{pythonpseudo}
system.appr
\end{pythonpseudo}
The \pyin{appr} object is created using
\begin{pythonpseudo}
appr = Approach(sys)
\end{pythonpseudo}
where \pyin{sys} is any object, which has attributes \pyin{qd}, \pyin{leads}, \pyin{si}, and \pyin{funcp}. For more details and structure see the source code of \pyin{Approach} class contained in \pyin{qmeq.aprclass} module.

For a custom approach the user can define a custom class, which needs to be derived from \pyin{Approach}. For example, let us take the \texttt{Python} implementation of the \emph{Lindblad} approach contained in \pyin{qmeq.approach.lindblad} module and define a custom class from it:
\begin{pythonpseudo}
from qmeq.aprclass import Approach
# Makes factors used for generating Lindblad master equation kernel
from qmeq.approach.lindblad import generate_tLba
# Generates master equation kernel
from qmeq.approach.lindblad import generate_kern_lindblad
# Calculates currents
from qmeq.approach.lindblad import generate_current_lindblad
# Acts on given reduced density matrix with kernel
from qmeq.approach.lindblad import generate_vec_lindblad

class ApproachCustom(Approach):

    kerntype = 'custom_pyLindblad'
    generate_fct = staticmethod(generate_tLba)
    generate_kern = staticmethod(generate_kern_lindblad)
    generate_current = staticmethod(generate_current_lindblad)
    generate_vec = staticmethod(generate_vec_lindblad)
\end{pythonpseudo}
We note that here we use static methods in order to have the same behavior between \texttt{Python} and \texttt{Cython} implementations. Then this custom approach can be used by setting {\ttm Builder(..., kerntype=ApproachCustom)}. Here we used already existing implementation of the Lindblad approach, but the user can redefine, for example, the function \pyin{generate_tLba}, which generates particular matrix elements of the jump operators.

\subsection{Parallelization}
In the present version \texttt{QmeQ} 1.0 the modules, which generate the Hamiltonian, the Liouvillian, and calculate the current, natively do not support parallelization. However, on computers with many cores the \texttt{QmeQ} code still can exploit parallelization through \texttt{NumPy}, when calculating matrix products, the eigenvalues, or matrix inverse, which is the most time consuming for larger systems. For \texttt{NumPy} to have the best multithreading capabilities it should be linked to so-called \texttt{ATLAS}/\texttt{OpenBLAS}/\texttt{MKL} libraries. The usage of many threads can be achieved by setting an environment variable \pyin{OMP_NUM_THREADS}. In \texttt{Python} this is done in the following way (here we use $8$ threads):
\begin{python}
import os
os.environ['OMP_NUM_THREADS'] = '8'
\end{python}
The above method may not always work in all operating systems. It is also possible to change \pyin{OMP_NUM_THREADS} using the terminal of a particular operating system. For example, on \texttt{Windows} terminal use {\ttm set OMP\_NUM\_THREADS=8} and on \texttt{Unix}/\texttt{Linux}/\texttt{MacOS} terminal use \pyin{export OMP_NUM_THREADS=8}. Lastly, when the calculations are made for large set of parameters it is always possible to use very simple poor-man's parallelization, where for every parameter value a new \texttt{Python} instance is run. This works when every instance is not too much memory consuming.

\section{\label{sec:ssorb}Example 1: Spinful single orbital}

In this section we elaborate more on the spinful single orbital example (see Eq.~\eqref{hamOne} and \figurename~\ref{sorb}) and show how to calculate a stability diagram for such a quantum dot. Here we parameterize the spin-up and spin-down energies as $\varepsilon_{\uparrow}=V_{g}+\frac{B}{2}$, $\varepsilon_{\downarrow}=V_{g}-\frac{B}{2}$, where $V_{g}$ is the gate voltage and $B$ is the magnetic field (representing anomalous Zeeman splitting of spinful orbital). First, we import all the relevant prerequisites
\begin{python}
# Prerequisites
import matplotlib.pyplot as plt
import numpy as np
import qmeq
\end{python}
and we choose such values for parameters:
\begin{python}
# Quantum dot parameters
vgate, bfield, omega, U = 0.0, 0.0, 0.0, 20.0
# Lead parameters
vbias, temp, dband = 0.5, 1.0, 60.0
# Tunneling amplitudes
gam = 0.5
t0 = np.sqrt(gam/(2*np.pi))
\end{python}
Here the variable \pyin{gam} represents $\Gamma=2\pi \abs{\text{\ttm t}_{0}}^2$ with $\text{\ttm t}_{0}=\sqrt{\nu_{F}}t_{0}$ (see Section~\ref{sec:rou}). In $H_{\mathrm{one}}$ \eqref{hamOne} we have two single particle states $\lvert\up\rangle$ and $\lvert\down\rangle$, which we label by $0$ and $1$, respectively. We also have four energy integrated lead channels $L\uparrow$, $R\uparrow$, $L\downarrow$, $R\downarrow$, which we label $0$, $1$, $2$, $3$. This system is constructed and solved with the \emph{Pauli} master equation in the following way
\begin{python}
nsingle = 2
# 0 is up, 1 is down
hsingle = {(0,0): vgate+bfield/2,
           (1,1): vgate-bfield/2,
           (0,1): omega}

coulomb = {(0,1,1,0): U}

tleads = {(0,0): t0, # L, up   <-- up
          (1,0): t0, # R, up   <-- up
          (2,1): t0, # L, down <-- down
          (3,1): t0} # R, down <-- down
                     # lead label, lead spin <-- level spin

nleads = 4
#        L,up        R,up         L,down      R,down
mulst = {0: vbias/2, 1: -vbias/2, 2: vbias/2, 3: -vbias/2}
tlst =  {0: temp,    1: temp,     2: temp,    3: temp}
system = qmeq.Builder(nsingle, hsingle, coulomb,
                      nleads, tleads, mulst, tlst, dband,
                      kerntype='Pauli')
system.solve()
\end{python}
We can check the obtained current and energy current using
\begin{python}
print('Pauli current:')
print(system.current)
print(system.energy_current)
\end{python}
\begin{outsty}
Pauli current:
[ 0.0207255 -0.0207255  0.0207255 -0.0207255]
[  1.73557597e-09  -1.73557597e-09   1.73557597e-09  -1.73557597e-09]
\end{outsty}
The four entries correspond to current in the four lead channels $L\uparrow$, $R\uparrow$, $L\downarrow$, $R\downarrow$. Currents through the left lead and the right lead channels are conserved up to numerical errors:
\begin{python}
print('Current continuity:')
print(np.sum(system.current))
\end{python}

\begin{outsty}
Current continuity:
2.77555756156e-17
\end{outsty}
If we want to change the approximate approach we could redefine the system with \pyin{qmeq.Builder} by specifying the new \pyin{kerntype}. It is also possible just to change the value of \pyin{system.kerntype}:
\begin{python}
kernels = ['Redfield', '1vN', 'Lindblad', 'Pauli']
for kerntype in kernels:
    system.kerntype = kerntype
    system.solve()
    print(kerntype, ' current:')
    print(system.current)
\end{python}

\begin{outsty}
Redfield  current:
[ 0.0207255 -0.0207255  0.0207255 -0.0207255]
1vN  current:
[ 0.0207255 -0.0207255  0.0207255 -0.0207255]
Lindblad  current:
[ 0.0207255 -0.0207255  0.0207255 -0.0207255]
Pauli  current:
[ 0.0207255 -0.0207255  0.0207255 -0.0207255]
\end{outsty}
For the considered system, the \emph{Pauli}, \emph{Redfield}, \emph{1vN}, and \emph{Lindblad} methods all yield the same results as for given parameter values spin is a good quantum number and no coherences between $\ket{\up}$ and $\ket{\down}$ are developed.

In order to use the \emph{2vN} approach, we will rebuild the system, because the solution method is rather different.\footnote{The solution procedure is implemented by a {\ttmfoot Approach2vN} object and not a {\ttmfoot Approach} object, which is used for the first-order approaches.} We also have to specify an equidistant energy grid on which the \emph{2vN} calculations are performed (iterative solution of an integral equation):

\begin{python}
kpnt = np.power(2,12)
system2vN = qmeq.Builder(nsingle, hsingle, coulomb,
                         nleads, tleads, mulst, tlst, dband,
                         kerntype='2vN', kpnt=kpnt)
\end{python}
Let us solve the \emph{2vN} integral equations using 7 iterations:
\begin{python}
system2vN.solve(niter=7)
print('Particle current:')
print(system2vN.current)
print('Energy current:')
print(system2vN.energy_current)
\end{python}

\begin{outsty}
Particle current:
[ 0.01595739 -0.01595739  0.01595739 -0.01595739]
Energy current:
[ 0.00599524 -0.00599523  0.00599524 -0.00599523]
\end{outsty}
We see that the particle current is reduced and the energy current is increased considerably compared to the \emph{Pauli} result. This is related to the inclusion of broadening in the \emph{2vN} approach. We can check the current for every iteration to see if it has converged:
\begin{python}
for i in range(system2vN.niter+1):
    print(i, system2vN.iters[i].current)
\end{python}
\begin{outsty}
0 [ 0.01505524 -0.01505524  0.01505524 -0.01505524]
1 [ 0.01589457 -0.01589457  0.01589457 -0.01589457]
2 [ 0.01595427 -0.01595427  0.01595427 -0.01595427]
3 [ 0.01595720 -0.01595720  0.01595720 -0.01595720]
4 [ 0.01595738 -0.01595738  0.01595738 -0.01595738]
5 [ 0.01595739 -0.01595739  0.01595739 -0.01595739]
6 [ 0.01595739 -0.01595739  0.01595739 -0.01595739]
\end{outsty}

In experiments usually the bias $V$ and gate voltage $V_{g}$ are controlled. A contour plot of current or conductance in the $(V, V_{g})$ plane is called a stability diagram. Let us produce such stability diagram for our spinful single orbital quantum dot using the \emph{Pauli} master equation. Thus we define a function for calculation

\begin{python}
def stab_calc(system, bfield, vlst, vglst, dV=0.0001):
    vpnt, vgpnt = vlst.shape[0], vglst.shape[0]
    stab = np.zeros((vpnt, vgpnt))
    stab_cond = np.zeros((vpnt, vgpnt))
    #
    for j1 in range(vgpnt):
        system.change(hsingle={(0,0):vglst[j1]+bfield/2,
                               (1,1):vglst[j1]-bfield/2})
        system.solve(masterq=False)
        for j2 in range(vpnt):
            system.change(mulst={0: vlst[j2]/2, 1: -vlst[j2]/2,
                                 2: vlst[j2]/2, 3: -vlst[j2]/2})
            system.solve(qdq=False)
            stab[j1, j2] = (system.current[0]
                          + system.current[2])
            #
            system.add(mulst={0: dV/2, 1: -dV/2,
                              2: dV/2, 3: -dV/2})
            system.solve(qdq=False)
            stab_cond[j1, j2] = (system.current[0]
                               + system.current[2]
                               - stab[j1, j2])/dV
    return stab, stab_cond
\end{python}
and a function for plotting
\begin{python}
def stab_plot(stab_cond, vlst, vglst, U, gam, title):
    (xmin, xmax, ymin, ymax) = np.array([vglst[0], vglst[-1],
                                         vlst[0], vlst[-1]])/U
    fig = plt.figure(figsize=(6,4.2))
    p = plt.subplot(1, 1, 1)
    p.set_xlabel('$V_{g}/U$', fontsize=20)
    p.set_ylabel('$V/U$', fontsize=20)
    p.set_title(title, fontsize=20)
    p_im = plt.imshow(stab_cond.T, extent=[xmin, xmax, ymin, ymax],
                                   aspect='auto',
                                   origin='lower',
                                   cmap=plt.get_cmap('Spectral'))
    cbar = plt.colorbar(p_im)
    cbar.set_label('Conductance $\mathrm{d}I/\mathrm{d}V$', fontsize=20)
    plt.show()
\end{python}
In \pyin{stab_calc} we changed the single-particle Hamiltonian by calling the function \pyin{system.change} and specifying which matrix elements to change. The function \pyin{system.add} adds a value to a specified parameter. Also the option \pyin{masterq=False} in \pyin{system.solve} indicates just to diagonalise the quantum dot Hamiltonian, but not to solve the master equation. Similarly, the option \pyin{qdq=False} means that the quantum dot Hamiltonian is not diagonalized (it was already diagonalized previously) and just master equation is solved. Let us now produce the stability diagram

\begin{figure}[t]
\begin{center}
\includegraphics[width=0.45\textwidth]{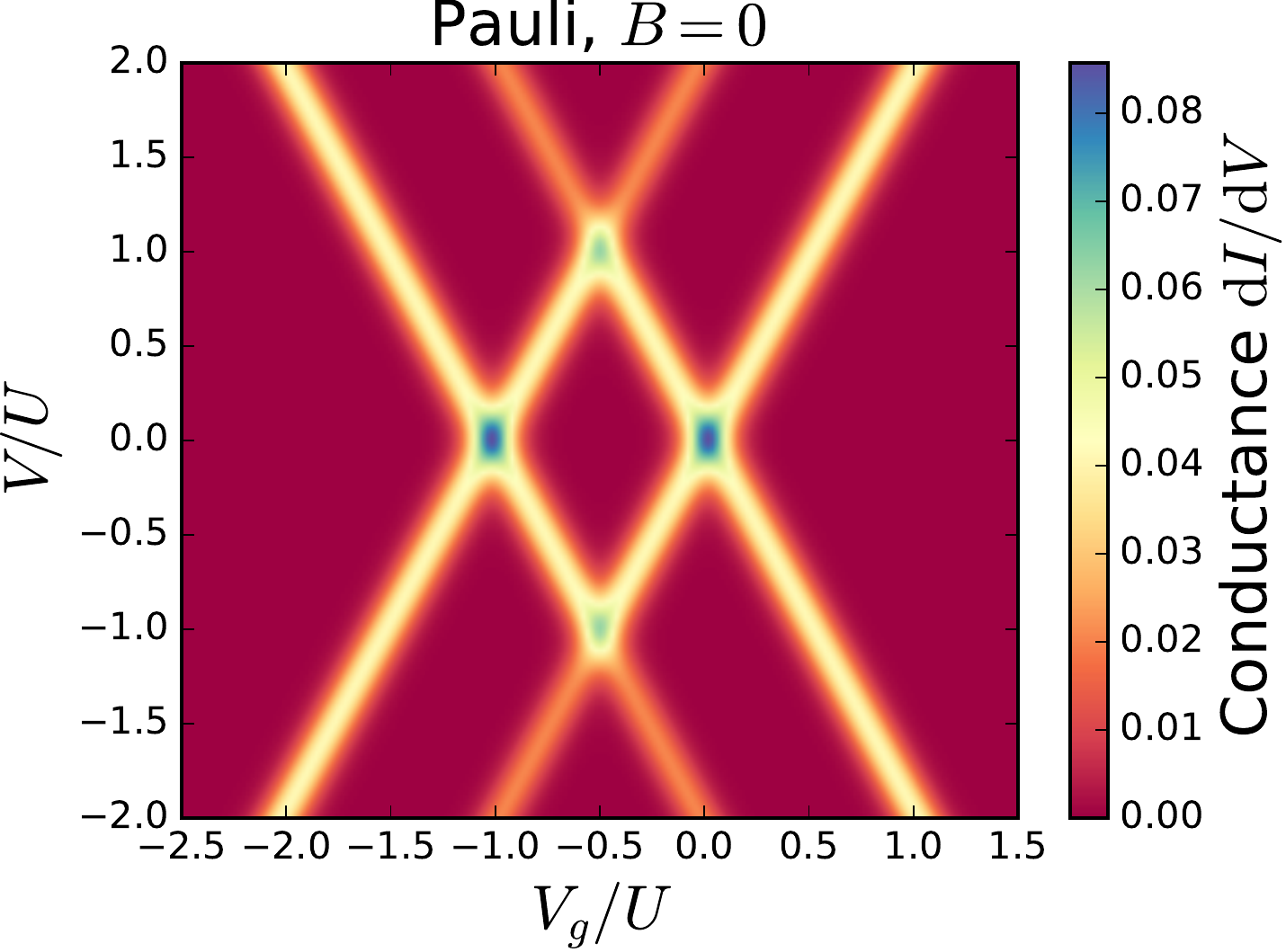}
\includegraphics[width=0.45\textwidth]{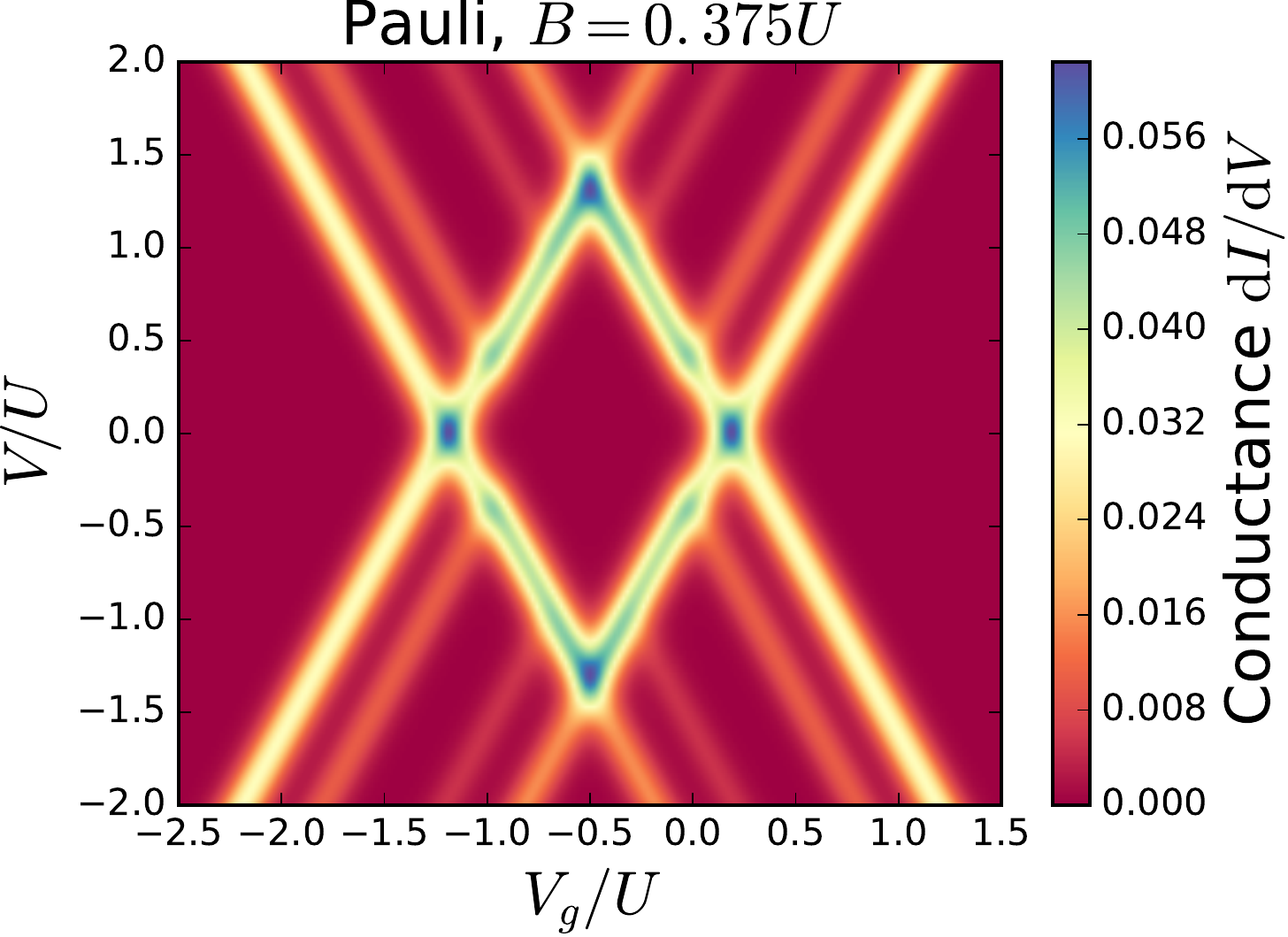}
\caption{\label{stab1_stab2} Stability diagrams calculated using the \emph{Pauli} approach for a spinful single orbital quantum dot without ($B=0$) and with ($B=0.375$) magnetic field. Other parameters are $\Gamma_{L}=\Gamma_{R}=T/2$, $U=20T$, $D=60T$.}
\end{center}
\end{figure}
%
%\begin{figure}[t]
%\begin{center}
%\includegraphics[width=0.9\textwidth]{stab2.pdf}
%\caption{\label{stab2} Stability diagram calculated using the \emph{Pauli} approach a for spinful single orbital quantum dot in the presence of magnetic %field $B=0.375U$. Other parameters are $\Gamma_{L}=\Gamma_{R}=T/2$, $U=20T$, $D=60T$.}
%\end{center}
%\end{figure}

\begin{python}
system.kerntype = 'Pauli'
vpnt, vgpnt = 201, 201
vlst = np.linspace(-2*U, 2*U, vpnt)
vglst = np.linspace(-2.5*U, 1.5*U, vgpnt)
stab, stab_cond = stab_calc(system, bfield, vlst, vglst)
stab_plot(stab_cond, vlst, vglst, U, gam, 'Pauli, $B=0$')
\end{python}
The result is shown in the left plot of \figurename~\ref{stab1_stab2}.

If a Zeeman splitting of the orbital is included, we obtain a stability diagram where the spin-split excited states can be seen:
\begin{python}
stab_b, stab_cond_b = stab_calc(system, 7.5, vlst, vglst)
stab_plot(stab_cond_b, vlst, vglst, U, gam, 'Pauli, $B=0.375U$')
\end{python}
which is shown in the right plot of \figurename~\ref{stab1_stab2}.

\begin{figure}[t]
\begin{center}
\includegraphics[width=0.5\textwidth]{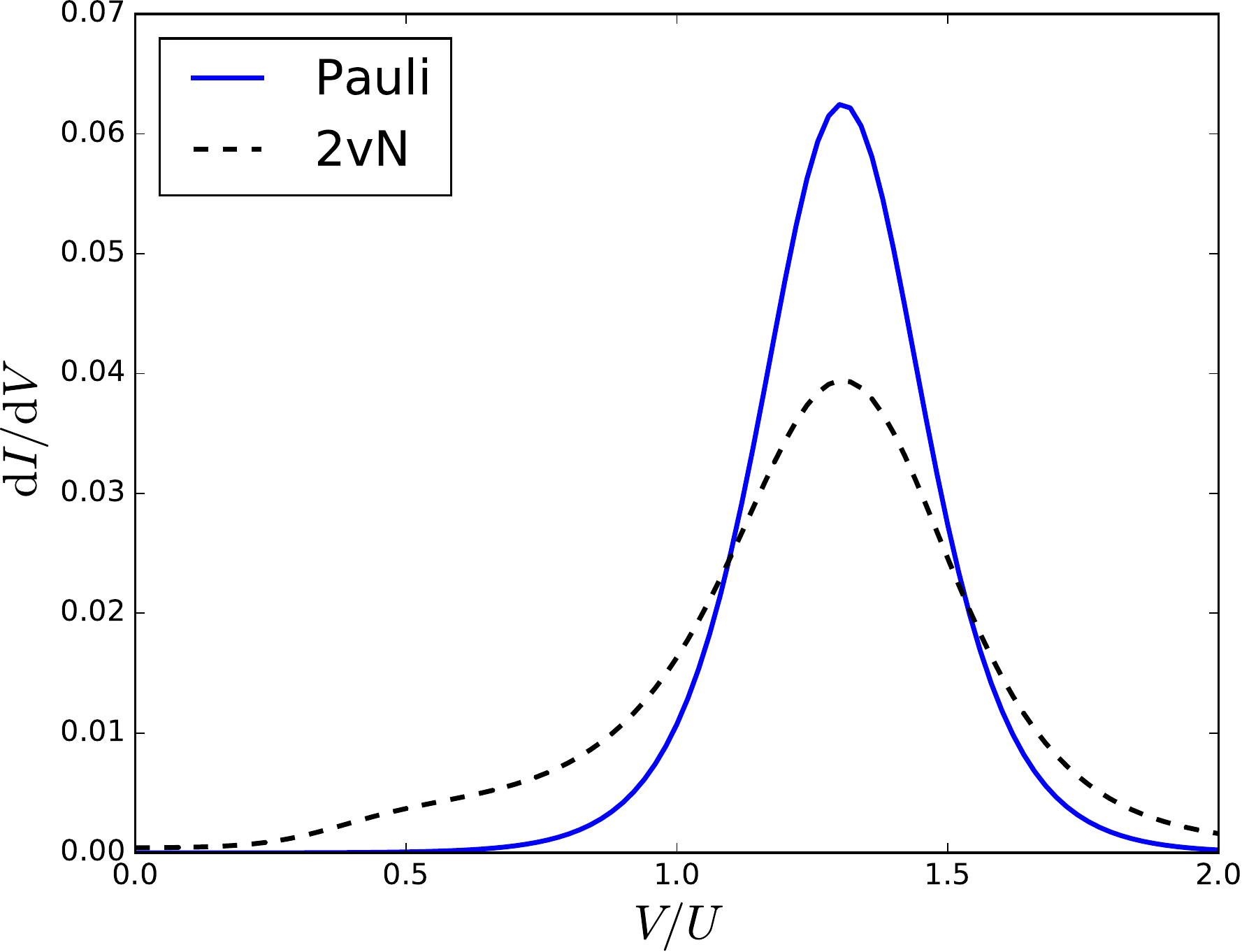}
\includegraphics[width=0.45\textwidth]{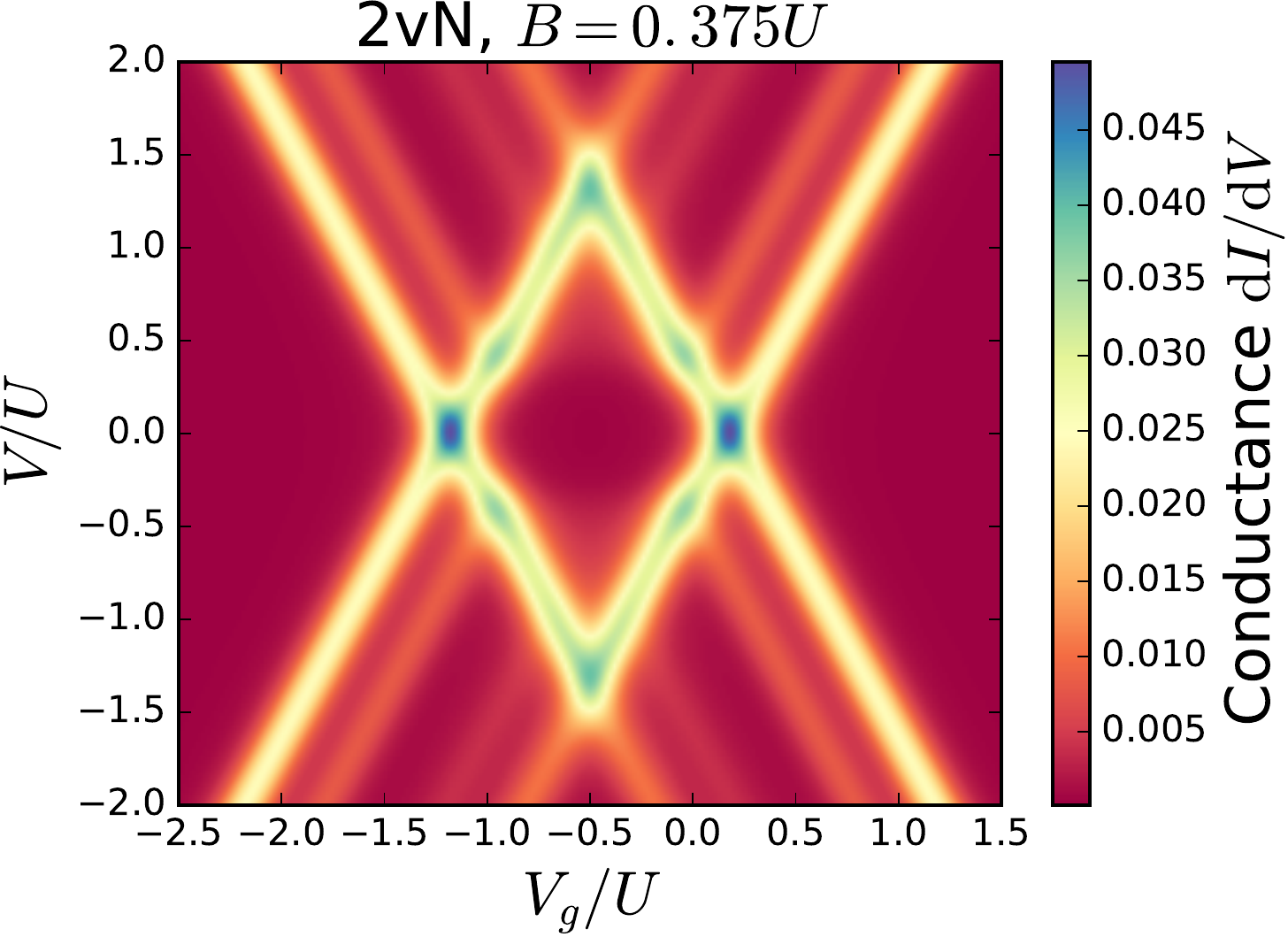}
\caption{Left plot: bias trace at the particle-hole symmetric point $V_{g}=-U/2$ for $B=0.375U$. Compared to the first-order \emph{Pauli} master equation the \emph{2vN} approach additionally yields a cotunneling conductance step appearing around $V=B$. Right plot: stability diagram calculated using the \emph{2vN} approach. Other parameters are $\Gamma_{L}=\Gamma_{R}=T/2$, $U=20T$, $D=60T$.}
\label{pauli2vN_stab2vN}
\end{center}
\end{figure}

Now we demonstrate the difference between the \emph{2vN} and the \emph{Pauli} approaches by considering the bias dependence of the conductance $\dif{I}/\dif{V}$ at the particle-hole symmetry point $V_{g}=-U/2$ for $B=0.375U$:
\begin{python}
def trace_vbias(system, vlst, vgate, bfield, dV=0.01, niter=7):
    vpnt = vlst.shape[0]
    trace = np.zeros(vpnt)
    #
    system.change(hsingle={(0,0): vgate+bfield/2,
                           (1,1): vgate-bfield/2})
    system.solve(masterq=False)
    #
    for j1 in range(vpnt):
        system.change(mulst={0: vlst[j1]/2, 1: -vlst[j1]/2,
                             2: vlst[j1]/2, 3: -vlst[j1]/2})
        system.solve(qdq=False, niter=niter)
        trace[j1] = (system.current[0]
                   + system.current[2])
        #
        system.add(mulst={0: dV/2, 1: -dV/2,
                          2: dV/2, 3: -dV/2})
        system.solve(qdq=False, niter=niter)
        trace[j1] = (system.current[0]
                   + system.current[2]
                   - trace[j1])/dV
    return trace

vpnt = 101
vlst = np.linspace(0, 2*U, vpnt)
trace_Pauli = trace_vbias(system, vlst, -U/2, 7.5)
trace_2vN = trace_vbias(system2vN, vlst, -U/2, 7.5)

fig = plt.figure()
p = plt.subplot(1, 1, 1)
p.set_xlabel('$V/U$', fontsize=20)
p.set_ylabel('$\mathrm{d}I/\mathrm{d}V$', fontsize=20)
plt.plot(vlst/U, trace_Pauli, label='Pauli',
                              color='blue',
                              lw=2)
plt.plot(vlst/U, trace_2vN, label='2vN',
                            color='black',
                            lw=2,
                            linestyle='--')
plt.legend(loc=2, fontsize=20)
plt.show()
\end{python}
The outcome is shown in the left plot of \figurename~\ref{pauli2vN_stab2vN}. We see that the \emph{2vN} approach captures the effect of cotunneling (conductance step appearing around $V=B$), while the first-order \emph{Pauli} master equation does not describe this effect.

Finally, we calculate the stability diagram for finite magnetic field using \emph{2vN} approach
\begin{python}
vpnt, vgpnt = 201, 201
vlst = np.linspace(-2*U, 2*U, vpnt)
vglst = np.linspace(-2.5*U, 1.5*U, vgpnt)
stab_b_2vN, stab_cond_b_2vN = stab_calc(system2vN, 7.5, vlst, vglst)
stab_plot(stab_cond_b_2vN, vlst, vglst, U, gam, '2vN, $B=0.375U$')
\end{python}
which is shown in the right plot of \figurename~\ref{pauli2vN_stab2vN}.

\section{\label{sec:sdqdot}Example 2: Spinless double quantum dot}

\begin{figure}[t]
\begin{center}
\includegraphics[width=0.5\textwidth]{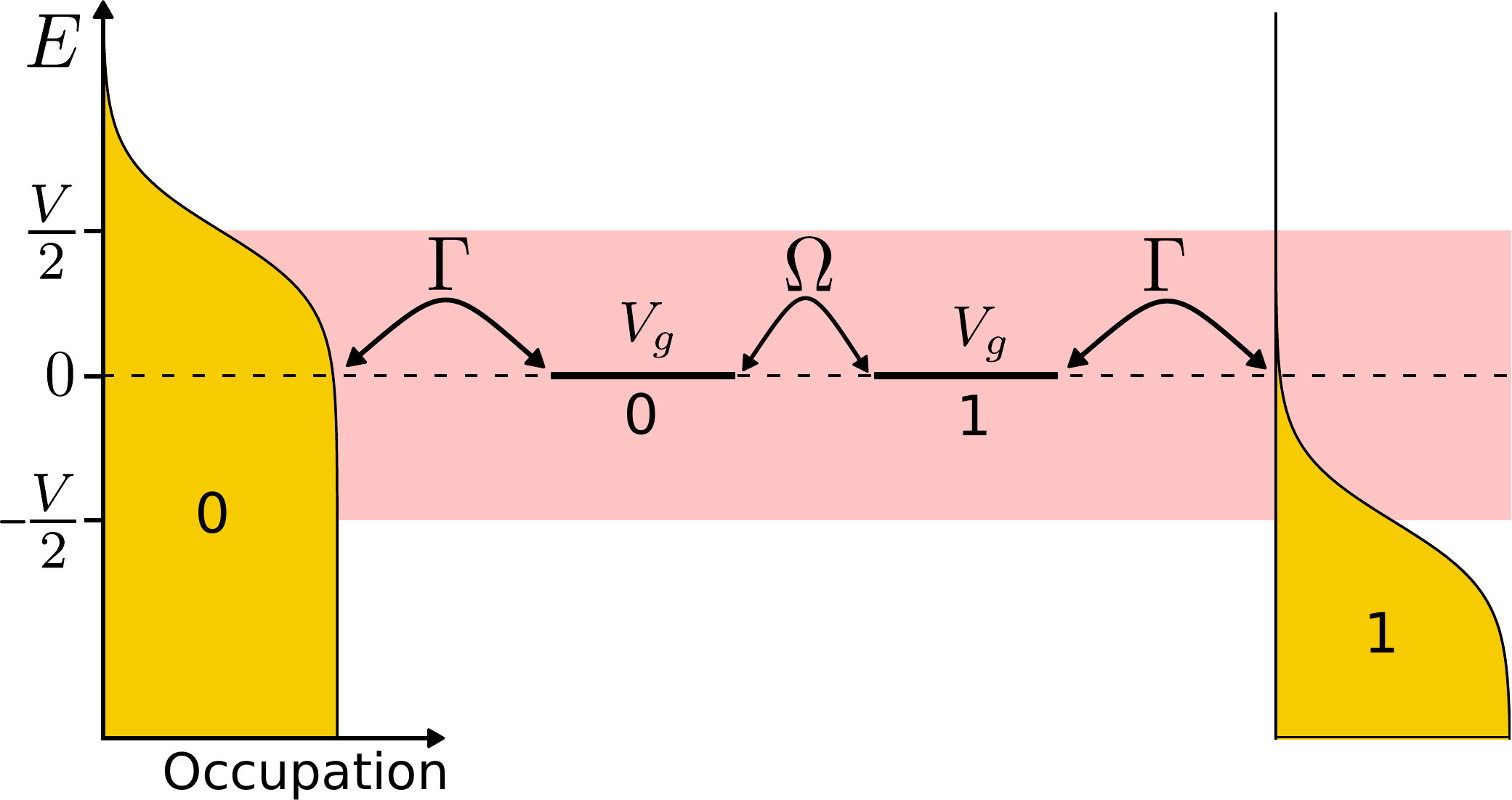}
\caption{Spinless double quantum dot structure. The energies of the dot states are shifted by a gate voltage $V_g$. Both dots are coupled to each other ($\Omega$) and to one lead each ($\Gamma$). The two leads have an applied bias voltage $V$, which gives a particle current $I$. }
\label{ddot}
\end{center}
\end{figure}

Our second example is spinless serial double quantum dot described by the Hamiltonian (see \figurename~\ref{ddot}):
\begin{equation}\label{hamTwo}
H_{\mr{two}}=\sum_{k;\ell=L,R}\ve_{\ell k}^{\phantom{\dag}}\cd_{\ell k}\can_{\ell k}
+\sum_{k}(t\dd_{l}\can_{L k}+t\dd_{r}\can_{R k}+\mr{H.c.})
+V_g(\dd_{l}\dan_{l} + \dd_{r}\dan_{r})+(\Omega\dd_{l}\dan_{r} +\mr{H.c.})
+U\dd_{l}\dd_{r}\dan_{r}\dan_{l}.
\end{equation}
As before we setup the system using \pyin{Builder}:
\begin{python}
# Prerequisites
import matplotlib.pyplot as plt
import numpy as np
import qmeq

# Quantum dot parameters
vgate, omega, U = 0.0, 2.0, 5.0
# Lead parameters
vbias, temp, dband = 0.5, 2.0, 60.0
# Tunneling amplitudes
gam = 1.0
t0 = np.sqrt(gam/(2*np.pi))

nsingle = 2

hsingle = {(0,0): vgate,
           (1,1): vgate,
           (0,1): omega}
coulomb = {(0,1,1,0): U}

tleads = {(0,0): t0, # L <-- l
          (1,1): t0} # R <-- r

nleads = 2
#        L           R
mulst = {0: vbias/2, 1: -vbias/2}
tlst =  {0: temp,    1: temp}

system = qmeq.Builder(nsingle, hsingle, coulomb,
                      nleads, tleads, mulst, tlst, dband)
\end{python}
We are interested in how the current depends the gate voltage $V_{g}$ and the hybridisation between the dots $\Omega$, which we determine using the function:
\begin{python}
def omega_vg(system, olst, vglst):
    opnt, vgpnt = olst.shape[0], vglst.shape[0]
    mtr = np.zeros((opnt, vgpnt), dtype=float)
    for j1 in range(opnt):
        system.change(hsingle={(0,1):olst[j1]})
        for j2 in range(vgpnt):
            system.change(hsingle={(0,0):vglst[j2],
                                   (1,1):vglst[j2]})
            system.solve()
            mtr[j1, j2] = system.current[0]
    return mtr
\end{python}
The plotting will be performed using:

\begin{figure}[t]
\begin{center}
\includegraphics[width=0.9\textwidth]{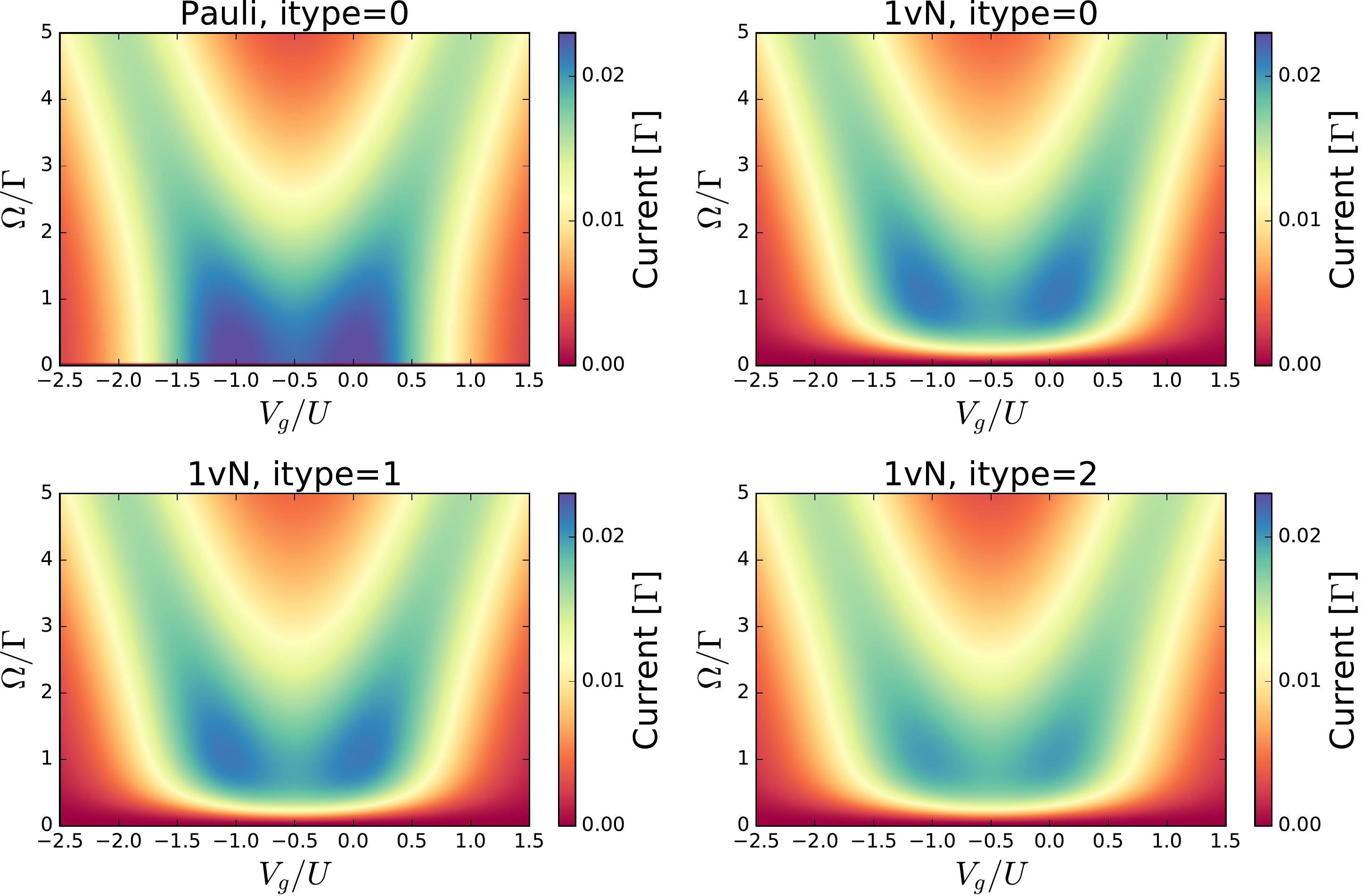}
\caption{\label{o_vg} The dependence of the current on $V_{g}$ and $\Omega$ for a spinless double quantum dot calculated using the \emph{Pauli} and the \emph{1vN} approaches. The values of the parameters are $\Gamma=T/2$, $U=5T/2$, $D=30T$. }
\end{center}
\end{figure}

\begin{python}
def plot_omega_vg(mtr, olst, vglst, U, gam, kerntype, itype, num):
    (xmin, xmax, ymin, ymax) = (vglst[0]/U, vglst[-1]/U,
                                olst[0]/gam, olst[-1]/gam)
    p = plt.subplot(2, 2, num)
    p.set_xlabel('$V_{g}/\Gamma$', fontsize=20)
    p.set_ylabel('$\Omega/\Gamma$', fontsize=20)
    p.set_title(kerntype+', itype='+str(itype), fontsize=20)
    p_im = plt.imshow(mtr/gam, extent=[xmin, xmax, ymin, ymax],
                               vmin=0, vmax= 0.023,
                               aspect='auto',
                               origin='lower',
                               cmap=plt.get_cmap('Spectral'))
    cbar = plt.colorbar(p_im)
    cbar.set_label('Current [$\Gamma$]', fontsize=20)
    cbar.set_ticks(np.linspace(0.0, 0.03, 4))
    plt.tight_layout()
\end{python}
Here, for the calculations we use the \emph{Pauli} approach and the \emph{1vN} approach, with three different values $0$, $1$, and $2$ for \pyin{Builder}'s optional argument \pyin{itype}. The value of \pyin{itype} determines how the integral of Eq.~\eqref{ooxi} is treated. In \texttt{QmeQ} for \pyin{itype=2}, the principal part $\mc{P}$ integrals are neglected, for \pyin{itype=1} the integrals are approximated using a digamma function \cite{AbramowitzBook1972}, and for \pyin{itype=0} a full calculation is performed using the \texttt{SciPy} implementation of the \texttt{QUADPACK} routine \texttt{DQAWC}. Now we calculate the current dependence on $V_{g}$ and $\Omega$ using the \emph{Pauli} and the \emph{1vN} methods:
\begin{python}
opnt, vgpnt = 201, 201
olst = np.linspace(0., 5., opnt)
vglst = np.linspace(-10., 10., vgpnt) - U/2

params = [['Pauli', 0, 1],
          ['1vN', 0, 2],
          ['1vN', 1, 3],
          ['1vN', 2, 4]]

fig = plt.figure(figsize=(12,8))

for kerntype, itype, num in params:
    system.kerntype = kerntype
    system.itype = itype
    mtr = omega_vg(system, olst, vglst)
    plot_omega_vg(mtr, olst, vglst, U, gam,
                  kerntype, itype, num)

plt.show()
\end{python}
The outcome of the calculation is shown in \figurename~\ref{o_vg}. We see that the \emph{Pauli} master equation gives the wrong current for $\Omega<\Gamma$, where the role of coherences is relevant and is correctly captured in \emph{1vN} approach (the \emph{Redfield}, \emph{Lindblad}, and \emph{2vN} approaches give similar results for the considered parameter values).

\section{\label{sec:sstd}Example 3: Spinful serial triple-dot structure}
\begin{figure}[t]
\begin{center}
\includegraphics[width=0.5\textwidth]{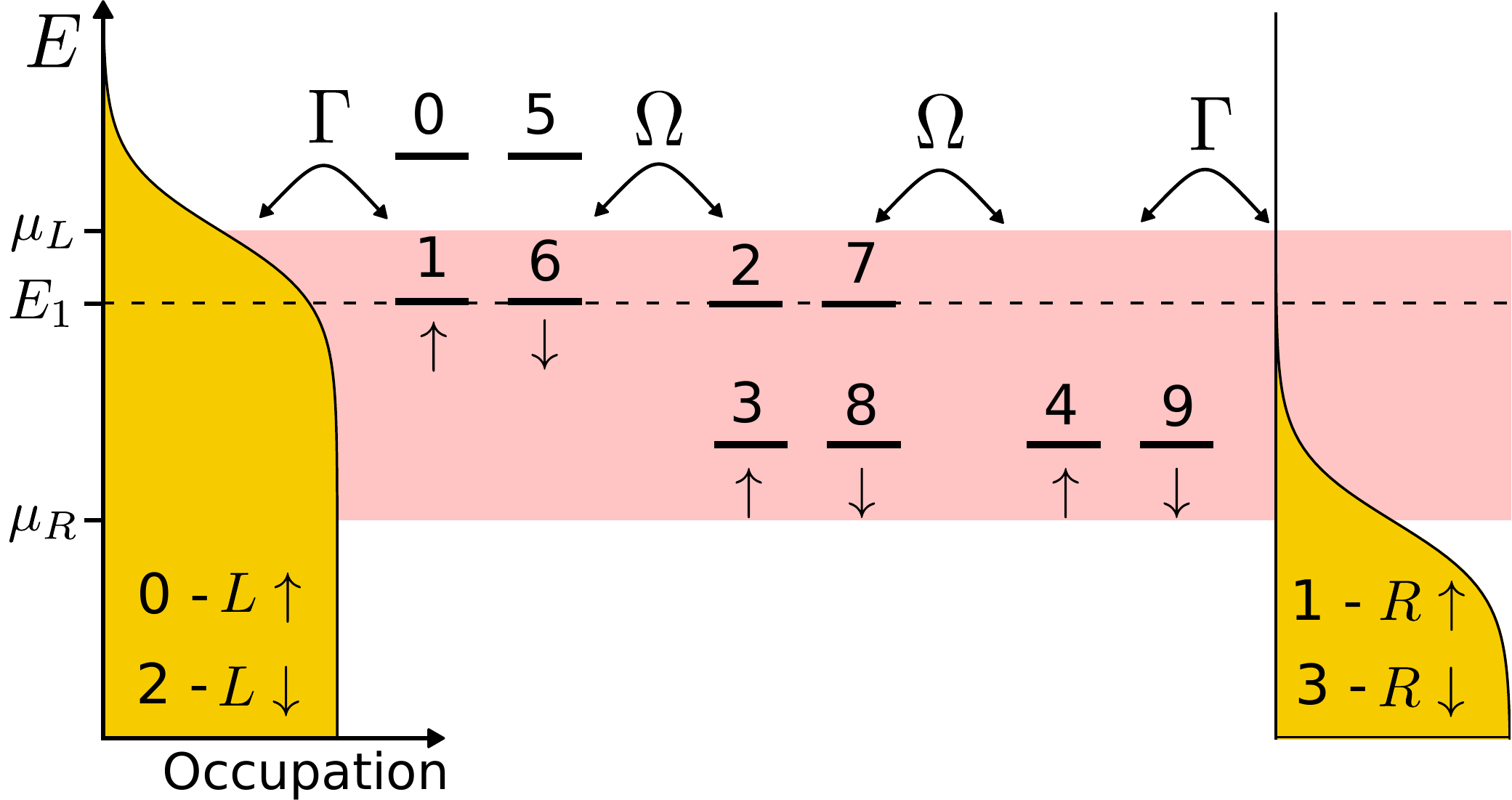}
\caption{A spinful serial triple-dot structure studied in Ref.~\cite{GoldozianSciRep2016}. It has $10$ single-particle states and $1024$ many-body states in total.}
\label{stdot}
\end{center}
\end{figure}

As a last example, we consider a system with many-degrees of freedom in the quantum dot region: the spinful serial triple-dot structure shown in \figurename~\ref{stdot}. The parameters for the model are summarized in \tablename~\ref{TableParameters}. In Ref.~\cite{GoldozianSciRep2016} the effect of different kinds of Coulomb matrix elements was considered in such a system. For the purposes of an example we here restrict ourselves to the dipole-dipole scattering of electrons between different quantum dots ($U_{sc}$). This term is responsible for the Auger process and is crucial for the current flow in the system.

Now we setup the system:
\begin{python}
# Prerequisites
import matplotlib.pyplot as plt
import numpy as np
import qmeq

# Lead and tunneling parameters
mul, mur = 50.0, 10.0
gam, temp, dband = 0.1, 1.0, np.power(10.0, 4)

tl, tr = np.sqrt(gam/(2*np.pi)), np.sqrt(gam/(2*np.pi))
tleads = {(0,0): +tl, (0,1): +tl, (1,4): -tr,
          (2,5): +tl, (2,6): +tl, (3,9): -tr}

nleads = 4
mulst = {0: mul,   1: mur,   2: mul,   3: mur}
tlst  = {0: temp,  1: temp,  2: temp,  3: temp}

# Quantum dot single-particle Hamiltonian
nsingle = 10
e0, e1, e2, e3, e4 = 60, 40, 38, 20, 20
o02, o03, o12, o13, o24, o34 = 0.2, 0.1, 0.1, -0.05, 0.2, 0.1

# Spin up Hamiltonian
hsingle0 = np.array([[e0,  0,   o02, o03, 0],
                     [0,   e1,  o12, o13, 0],
                     [o02, o12, e2,  0,   o24],
                     [o03, o13, 0,   e3,  o34],
                     [0,   0,   o24, o34, e4]])
# Augment the Hamiltonian to have spin up and down
hsingle = np.kron(np.eye(2), hsingle0)

# Coulomb matrix elements
usc = -0.2
coulomb = {(0,2,3,1):usc, (0,3,2,1):usc, (0,7,8,1):usc, (0,8,7,1):usc,
           (1,2,3,0):usc, (1,3,2,0):usc, (1,7,8,0):usc, (1,8,7,0):usc,
           (2,5,6,3):usc, (2,6,5,3):usc, (3,5,6,2):usc, (3,6,5,2):usc,
           (5,7,8,6):usc, (5,8,7,6):usc, (6,7,8,5):usc, (6,8,7,5):usc}

system = qmeq.Builder(nsingle, hsingle, coulomb,
                      nleads, tleads, mulst, tlst, dband,
                      indexing='ssq', kerntype='Pauli', itype=2)
\end{python}
Here we use the spin symmetry of the problem by setting \pyin{indexing='ssq'}, which is very relevant for reducing the Liouvillian size and memory consumption in the \emph{1vN} approach calculations. Also we neglect the principal part integrals by setting \pyin{itype=2}. Additionally, we have set up the single-particle Hamiltonian using a NumPy array.\footnote{For more information on allowed input types in \texttt{QmeQ} see the \href{https://github.com/gedaskir/qmeq-examples/tree/master/appendix/00_types.ipynb}{00\_types.ipynb} notebook.}

\begin{table}[t]
\begin{center}
\begin{tabular}{l|l|l}
  \hline
     $E_{0}=E_{5}=60$ & $U_{sc}=-0.2$ & $\Omega_{02}=\Omega_{57}=0.2$ \\ %\hline
     $E_{1}=E_{6}=40$ & $\Gamma=0.1$  & $\Omega_{03}=\Omega_{58}=0.1$ \\ %\hline
     $E_{2}=E_{7}=38$ & $\mu_{L}=50$  & $\Omega_{12}=\Omega_{67}=0.1$ \\ %\hline
     $E_{3}=E_{8}=20$ & $\mu_{R}=10$  & $\Omega_{13}=\Omega_{68}=-0.05$\\ %\hline
     $E_{4}=E_{9}=20$ & $T=1$         & $\Omega_{24}=\Omega_{79}=0.2$ \\ %\hline
                      & $D=10^4$      & $\Omega_{34}=\Omega_{89}=0.1$ \\
  \hline
\end{tabular}
\end{center}
\caption{\label{TableParameters} Parameters used in the calculations of transport through a triple dot. }
\end{table}

\begin{figure}[t]
\begin{center}
\includegraphics[width=0.5\textwidth]{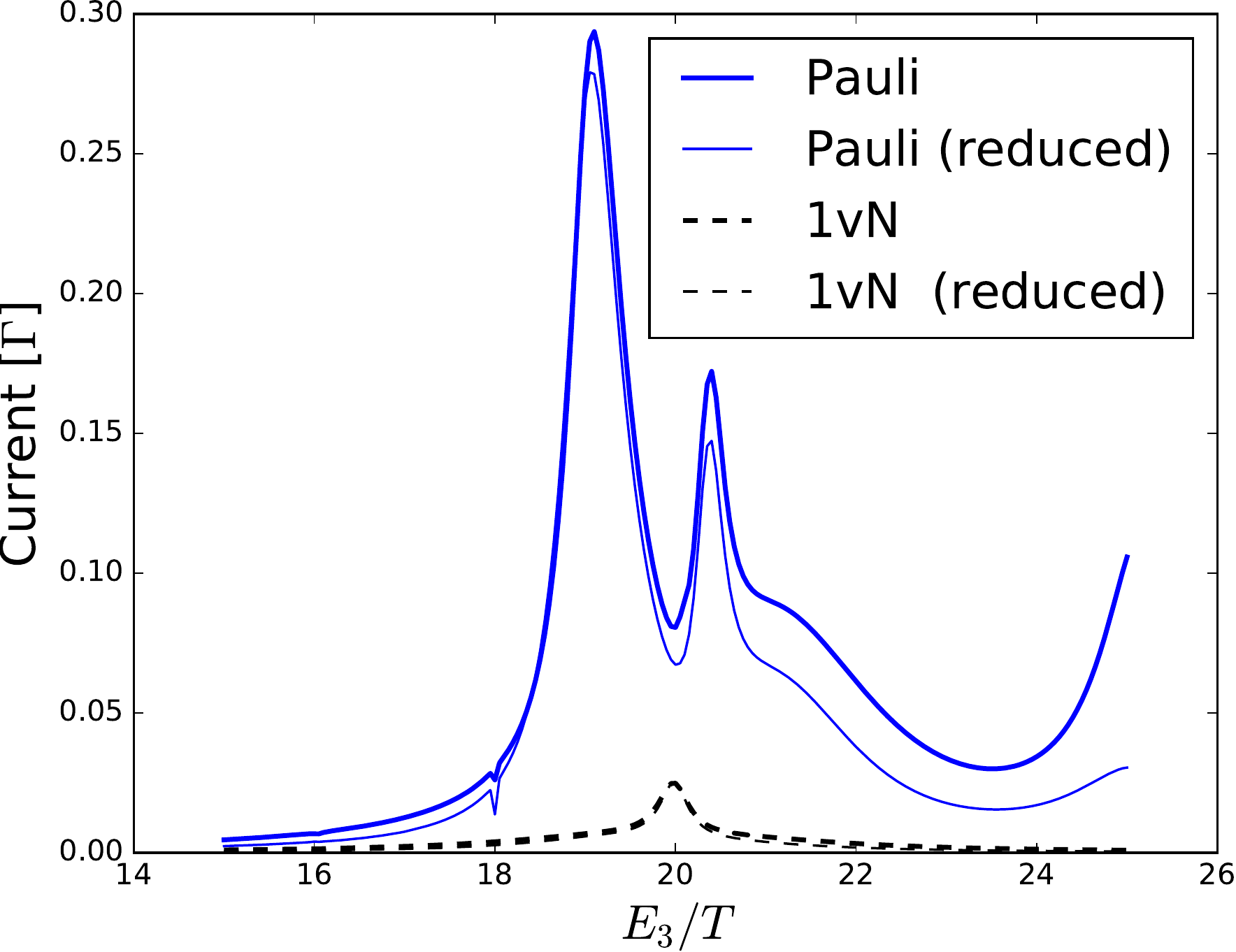}
\caption{The dependence of the current on the level position $E_{3}=E_{8}$ for the serial triple-dot structure. The thin curves ``Pauli (reduced)'' and ``1vN (reduced)'' correspond to the calculation when the many-body states with energy $\Delta{E}=150T$ above the ground state are removed. For the \emph{1vN} calculation this removal considerably reduces the Liouvillian size and still gives reasonable results.}
\label{pauli1vN}
\end{center}
\end{figure}

Assume we are interested in how the current depends on the level position $E_{3}=E_{8}$, which we calculate using:
\begin{python}
def trace_e3(system, e3lst, removeq=False, dE=150.0):
    e3pnt = e3lst.shape[0]
    trace = np.zeros(e3pnt)
    system.use_all_states()
    for j1 in range(e3pnt):
        system.change({(3, 3): e3lst[j1],
                       (8, 8): e3lst[j1]})
        system.solve(masterq=False)
        if removeq:
            system.remove_states(dE)
        system.solve(qdq=False)
        trace[j1] = sum(system.current[np.ix_([0,2])])
    return trace

e3pnt = 201
e3lst = np.linspace(15., 25., e3pnt)

system.kerntype = 'Pauli'
tr_e3_Pauli = trace_e3(system, e3lst)
tr_e3_Pauli_rm = trace_e3(system, e3lst, removeq=True)
system.kerntype = '1vN'
tr_e3_1vN = trace_e3(system, e3lst)
tr_e3_1vN_rm = trace_e3(system, e3lst, removeq=True)
fig = plt.figure()
p = plt.subplot(1, 1, 1)
p.set_xlabel('$E_{3}/T$', fontsize=20)
p.set_ylabel('Current [$\Gamma$]', fontsize=20)
plt.plot(e3lst/temp, tr_e3_Pauli/gam,
         label='Pauli', color='blue', lw=2)
plt.plot(e3lst/temp, tr_e3_Pauli_rm/gam,
         label='Pauli (reduced)', color='blue', lw=1)
plt.plot(e3lst/temp, tr_e3_1vN/gam,
         label='1vN', color='black', lw=2, linestyle='--')
plt.plot(e3lst/temp, tr_e3_1vN_rm/gam,
         label='1vN  (reduced)', color='black', lw=1, linestyle='--')
plt.legend(loc=1, fontsize=20)
plt.show()
\end{python}
The result is shown in \figurename~\ref{pauli1vN}. We see that the current is considerably reduced in the \emph{1vN} approach compared to the first-order \emph{Pauli} master equation. This reduction is the effect of the coherences developing between the many-body states which have energy differences smaller than the tunneling rate $\Gamma$, similarly as discussed in the simple double-dot example in Section~\ref{sec:sdqdot}.

An important thing to note is how we used the spin symmetry (\pyin{indexing='ssq'}).\footnote{For more information on usage of symmetries in \texttt{QmeQ} see the \href{https://github.com/gedaskir/qmeq-examples/tree/master/appendix/01_symmetries.ipynb}{01\_symmetries.ipynb} notebook.} In order to be able to use the spin symmetry in the problem the single particle Hamiltonian, tunneling amplitudes, and the lead parameters need to be set up in a particular way. Let us say we have $n=2m$ quantum dot single particle states counting with spin. Then we use the convention that the states with spin up have the labels $0\ldots m-1$ and the states with spin down have the labels $m\ldots n-1$. Also there should be no coupling between the up and down states to obtain correct results. Additionally, when the spin degeneracy is present and we want to have a number $m_{\alpha}$ of physical leads, then we need to specify parameters for $n_{\alpha}=2m_{\alpha}$ number of leads. For example, if there are source and drain leads with chemical potentials $\mu_{L}$ and $\mu_{R}$, then we need to specify the following dictionary for chemical potentials:
\begin{pythonpseudo}
mulst = {0: $\mu_{L}$, 1: $\mu_{R}$,  # spin up
         2: $\mu_{L}$, 3: $\mu_{R}$}  # spin down
\end{pythonpseudo}

Now we shortly discuss how the number of $\Phi^{[0]}$ matrix elements is reduced using the spin symmetry. The many-body eigenstates $\ket{b}=\ket{N,S,M,i}$ are classified by the number of particles $N$, the total spin value $S$, and the spin-projection value $M$, and $i$ denotes some other quantum numbers. In the stationary state, due to selection rules the only non-vanishing reduced density matrix $\Phi^{[0]}_{bb'}$ elements are between the states $\ket{b}=\ket{N,S,M,i}$ and $\ket{b'}=\ket{N,S,M,i'}$ for arbitrary $i$ and $i'$. Additionally, all the matrix elements are equal when the states have the same $N$, $S$, $i$, and $i'$ quantum numbers, e.g., $\Phi^{[0]}_{(N,S,M,i),(N,S,M,i')}=\Phi^{[0]}_{(N,S,M',i),(N,S,M',i')}$.

Lastly, to manually specify all the matrix elements related by spin symmetry can be tedious and error prone. For this reason it is possible to set up the system by giving only the values for spin-up direction and using optional argument \pyin{symmetry='spin'} in the \pyin{Builder}. This will automatically augment the parameters to be consistent with the spin symmetry and will use \pyin{indexing='ssq'} if it is supported by the implementation of the approach. Then for the serial triple dot considered above we can write:
\begin{python}
tleads = {(0,0): +tl, (0,1): +tl, (1,4): -tr}

mulst = {0: mul,   1: mur}
tlst  = {0: temp,  1: temp}

coulomb = {(0,2,3,1):usc, (0,3,2,1):usc,
           (1,3,2,0):usc, (1,2,3,0):usc}

system = qmeq.Builder(nsingle, hsingle0, coulomb,
                      nleads, tleads, mulst, tlst, dband,
                      kerntype='Pauli', itype=2,
                      symmetry='spin')
\end{python}
We also note that if a direct interaction element like $U\dd_{0}\dd_{5}\dan_{5}\dan_{0}=U\dd_{0\up}\dd_{0\down}\dan_{0\down}\dan_{0\up}$ is needed and \pyin{symmetry='spin'} is used, then the user needs to specify $U\dd_{0}\dd_{0}\dan_{0}\dan_{0}$ in \pyin{coulomb}.

\section{Computational cost}

In this section we discuss the memory requirements for different approaches and computation time for examples considered in this paper. For $n$ single-particle states we have
\begin{enumerate}
\item[\textbf{*}] $N=2^{n}$ many-body states,
\item[\textbf{*}] $N_{0}=\frac{(2n)!}{(n!)^2}$ reduced density matrix $\Phi^{[0]}$ elements,
\item[\textbf{*}] $N_{1}=\frac{n}{n+1}N_{0}$ current amplitude $\Phi^{[1]}$ elements.
\end{enumerate}
Here the estimate of $\Phi^{[0]}$ is done assuming that there are no coherences between the many-body states containing different number of particles, for example, $\Phi^{[0]}_{ab}=0$, $\Phi^{[0]}_{ac}=0$, and etc. When the Hamiltonian conserves the total number of particles such off-diagonal elements decay rapidly due to superselection rules and are not present in the stationary state \cite{ZurekPRD1982}. Additionally, in the considered approaches when the reduced density matrix is diagonal in some conserved quantity (e.~g., charge) at some point in time (e.~g., in the infinite past), it will remain diagonal for all times. The estimate of $\Phi^{[1]}$ is done by counting only the matrix elements of the form $\Phi^{[1]}_{cb}$.

\begin{figure}[t]
\begin{center}
\includegraphics[width=0.6\textwidth]{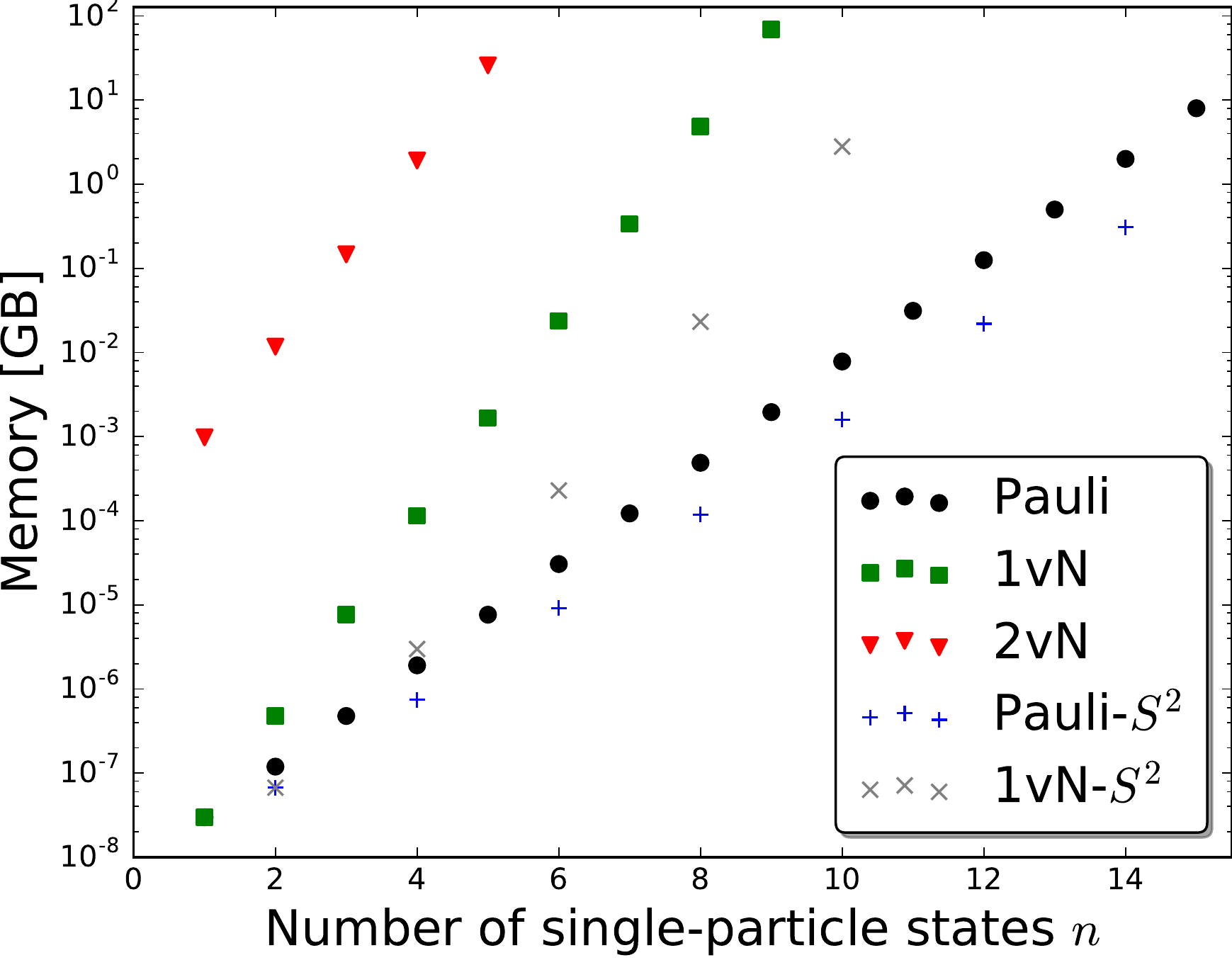}
\caption{Memory requirement dependence on the number of single-particle states $n$ when all many-body states are kept for the calculation. To make an estimate for the \emph{2vN} approach we used $N_{\alpha}=2$, $N_{k}=2^{12}$. Pauli-$S^2$ and 1vN-$S^2$ correspond to memory requirement, when the spin symmetry is used.}
\label{memreq}
\end{center}
\end{figure}

To store the \emph{kernel} for the first-order approaches or the solution (with related quantities) for the \emph{2vN} approach we need:
\begin{enumerate}
\item[\textbf{*}] $64\times N^2$ bits for \emph{Pauli},
\item[\textbf{*}] $64\times (2N_{0}-N)^2$ bits for \emph{1vN}, \emph{Redfield}, \emph{Lindblad},
\item[\textbf{*}] $4\times 128\times N_{\alpha}N_{k}N_{0}N_{1}$ bits for \emph{2vN},
\end{enumerate}
where we assumed 64-bit floating point arithmetics. Here $N_{\alpha}$ is the number of lead channels and $N_{k}$ is the number of energy grid points. If all many-body states are kept the memory requirement depending on the number of single-particle states $n$ is shown in \figurename~\ref{memreq} (for the \emph{2vN} approach we used $N_{\alpha}=2$, $N_{k}=2^{12}$). For example, if $8$GB of memory is available then it is possible to solve a system with the following number of single-particle states: $15$ for \emph{Pauli}, $8$ for \emph{1vN}, $4$ for \emph{2vN}.\footnote{We note that for the first-order approaches it is possible to use matrix-free methods for solution by specifying an optional argument {\ttmfoot Builder(..., mfreeq=True)}. Then the memory requirement is much lower and is proportional to $N$ or $N_{0}$ for \emph{Pauli} or \emph{1vN}, \emph{Redfield}, \emph{Lindblad} approaches, respectively. However, the computation time increases considerably.}

The spinful serial triple-dot structure (see Section~\ref{sec:sstd}) with $n=10$ single-particle states is an example, which requires a lot of memory. With all the many-body states present the spin-symmetry of the problem allows to reduce the memory requirement from $1$TB to $3$GB for the \emph{1vN} approach. The calculation for a single $E_{3}$ point (see \figurename~\ref{pauli1vN}) takes 170 seconds on our computer based on 3.50GHz processor (\emph{Intel Xeon E3-1270v3}).
For the calculation, where the many-body states with energy $\Delta{E}=150T$ above the ground state are removed, the memory requirement is reduced to $0.1$GB and computation time is reduced to 4 seconds.
%For \emph{Pauli} master equation the memory requirement for the \emph{kernel} is low: $\sim 8$MB with no symmetry used and $\sim 2$MB with spin-symmetry used.

In the spinful single orbital example (see Section~\ref{sec:sstd}) the \emph{2vN} approach takes 24 seconds to calculate a single point of current on our computer with 3.50GHz processor. So it takes around 1.5 hour to produce the left plot of \figurename~\ref{pauli2vN_stab2vN} with $2\times101$ points of calculation and around 540 hours to produce the right plot of \figurename~\ref{pauli2vN_stab2vN} with $2\times80802$ points of calculation. We note that because of slow speed and large memory requirement for larger systems (see \figurename~\ref{memreq}) the \emph{2vN} approach is useful only for describing systems with small number of many-body states.

\section{\label{sec:rou}Remark on units}
The inputted tunneling amplitudes \pyin{t$_{\leadqn i}$} in \pyin{tleads} correspond to tunneling amplitudes $t_{\leadqn i}$ weighted by the density of states $\nu_{F}$ in the following way:
\begin{equation}
\text{\ttm t}_{\leadqn i}=\sqrt{\nu_{F}}t_{\leadqn i}.
\end{equation}
Then the tunneling rates simply become $\Gamma_{\leadqn i}=2\pi\abs{\text{\ttm t}_{\leadqn i}}^2$. When different lead channels can have different density of states $\nu_{\leadqn i}$, any difference can be absorbed into the tunneling amplitudes. Thus it was not necessary to specify $\nu_{F}$ in the given examples.

Throughout the calculations we have used $\hbar=1$ and considered particle currents instead of electrical currents. If we are interested in the case $\hbar\neq 1$, $\abs{e}\neq 1$ and in electrical currents with carriers having a charge $e$ (which can be $e<0$) we have to make the following changes in Figures~\ref{stab1_stab2}--\ref{pauli1vN}:
\begin{equation}
\begin{aligned}
V&\rightarrow eV,\\
V_{g}&\rightarrow eV_{g},\\
I \ [\Gamma]&\rightarrow I \ [e\Gamma/\hbar],\\
\frac{\dif{I}}{\dif{V}} \ [1]&\rightarrow \frac{\dif{I}}{\dif{V}} \ [e^2/\hbar].
\end{aligned}
\end{equation}
To get the conductance in units of $G_{0}=e^2/h$, we should multiply the $\dif{I}/\dif{V}$ plots of Figures~\ref{stab1_stab2} and \ref{pauli2vN_stab2vN} by $2\pi$.

\section{Conclusion}

We have presented an open-source \texttt{Python} package \texttt{QmeQ} for numerical modeling of stationary state transport through quantum dot devices, which can be described by the tunneling model \eqref{ham} with quantum dots having arbitrary Coulomb interactions. The main objective of the package is to provide numerical results based on various approximate master equation approaches. We have introduced and explained the features of the package in an example driven way. The version of the package described in this paper is \texttt{QmeQ} 1.0.

\section*{Acknowledgements}

We thank K.~M.~Seja, F.~A.~Damtie, B.~Goldozian, and K.~G.~L.~Pedersen for useful discussions. Financial support from the Swedish Research Council (VR) and NanoLund is gratefully acknowledged.

\appendix

\section{\label{App} Approximate master equations}

\begin{figure}[t]
\begin{center}
\includegraphics[width=0.6\textwidth]{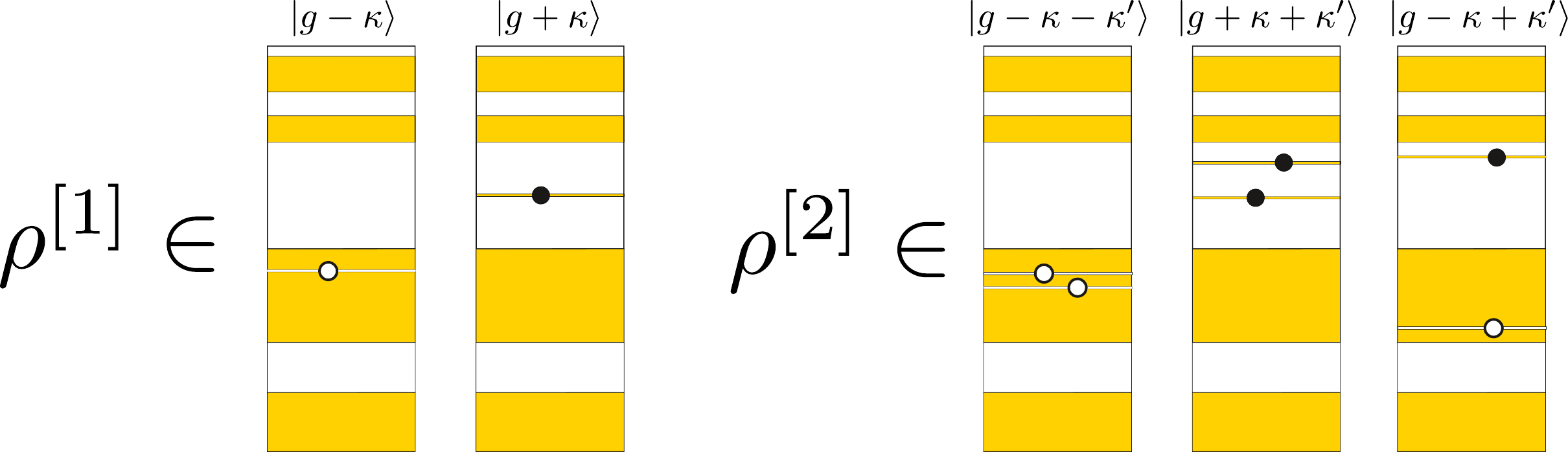}
\caption{\label{ehx} Classification of the density matrix elements by the number of electron/hole excitations in the leads. $\rho^{[1]}$ or $\rho^{[2]}$ shows cases when there is a single or two electron/hole excitations in the leads, respectively. Solid dots represent electrons and hollow dots represent holes.}
\end{center}
\end{figure}

In the Appendices we present the derivation of the approximate master equations implemented in \texttt{QmeQ} %so far
and discuss their properties. These are
the \textit{second-order von Neumann} (\ref{App:2vN}),
the \textit{first-order von Neumann} (\ref{App:1vN}),
the \textit{first-order Redfield} (\ref{App:Redfield}),
and
the \textit{Pauli} (\ref{App:Pauli})
approaches.
We also shortly describe a particular form of the \textit{Lindblad} equation in \ref{App:Lindblad}, which takes first-order processes into consideration. It is similar to the \emph{first-order Redfield} or \emph{first-order von Neumann} methods, but additionally preserves positivity of the reduced density matrix \cite{Kirsanskas2017Lindblad}. In \ref{App:Which} we give suggestions in which cases which approach to use.

We are interested in solving the von Neumann equation approximately. This equation describes the evolution of the density matrix $\rho$ of the whole system:
\begin{equation}\label{vneq}
\ii\frac{\pd}{\pd t}\rho=[H,\rho].
\end{equation}
Our derivation of the approaches is based on the hierarchical method, where we classify the density matrix elements by the number of electron/hole excitations in the leads (see \figurename~\ref{ehx}). Such a derivation was originally presented in Refs.~\cite{PedersenPRB2005,PedersenPHE2010}, but most of this Appendix is based on Refs.~\cite{GoldozianSciRep2016,KirsanskasPRB2016}. Here we derive the necessary equations to define the \emph{second-order von Neumann} approach. Lower-order approaches will be deductively obtained in succeeding sections by making additional approximations to these equations. Density matrix elements are defined and classified as
\begin{equation}
\rho_{ag,bg'}^{[n]}=\bra{ag}\rho\ket{bg'},
\end{equation}
where $\ket{bg}=\ket{b}\otimes\ket{g}$, with $\ket{b}$ denoting the eigenstate of the dot Hamiltonian~$H_{\mr{dot}}$~\eqref{hamQD} and $\ket{g}$ denoting the eigenstate of the lead Hamiltonian~$H_{\mr{leads}}$~\eqref{hamLeads}. Here the label $n$ provides the number of electron or hole excitations needed to transform $\ket{g}$ into $\ket{g'}$.
For example, we consider the matrix elements of the type
\begin{equation}
\begin{aligned}
&\rho_{bg,b'g}^{[0]}=\bra{bg}\rho\ket{b'g}, &&\rho_{dg-\kappa-\kappa',bg}^{[2]}=\bra{dg-\kappa-\kappa'}\rho\ket{bg},\\
&\rho_{bg-\kappa,ag}^{[1]}=\bra{bg-\kappa}\rho\ket{ag}, &&\rho_{bg-\kappa+\kappa',b'g}^{[2]}=\bra{bg-\kappa+\kappa'}\rho\ket{b'g}.
\end{aligned}
\end{equation}
Here we have introduced a composite index
\begin{equation}
\kappa\equiv k, \ \leadqn;
\end{equation}
and the following notation
\begin{equation}
\begin{aligned}
&\ket{bg+\kappa}=\ket{b}\otimes\cd_{\kappa}\ket{g}, &&\ket{dg-\kappa-\kappa'}=\ket{d}\otimes\can_{\kappa'}\can_{\kappa}\ket{g},\\
&\ket{bg-\kappa}=\ket{b}\otimes\can_{\kappa}\ket{g}, &&\ket{bg-\kappa+\kappa'}=\ket{b}\otimes\cd_{\kappa'}\can_{\kappa}\ket{g}.\\
\end{aligned}
\end{equation}
By neglecting all the density matrix elements with more than two electron or hole excitations $n>2$ from Eq.~\eqref{vneq} we obtain the equations
\begin{fleqn}
\begin{equation}\label{rheq0}
\begin{aligned}
\ii\frac{\pd}{\pd{t}}\rho_{bg,b'g}^{[0]}=(E_{b}-E_{b'})\rho_{bg,b'g}^{[0]}
&+\shs{\sum_{a_1,\kappa_1}}T_{ba_1,\kappa_1}\rho_{a_1g+\kappa_1,b'g}^{[1]}(-1)^{N_{a_1}}
+\shs{\sum_{c_1,\kappa_1}}T_{bc_1,\kappa_1}\rho_{c_1g-\kappa_1,b'g}^{[1]}(-1)^{N_{b}}\\
&-\shs{\sum_{c_1,\kappa_1}}\rho_{bg,c_1g-\kappa_1}^{[1]}(-1)^{N_{b'}}T_{c_1b',\kappa_1}
-\shs{\sum_{a_1,\kappa_1}}\rho_{bg,a_1g+\kappa_1}^{[1]}(-1)^{N_{a_1}}T_{a_1b',\kappa_1},
\end{aligned}
\end{equation}
\begin{equation}\label{rheq1}
\begin{aligned}
\ii\frac{\pd}{\pd{t}}\rho_{cg-\kappa,bg}^{[1]}
=(E_{c}-\leadE{\kappa}-E_{b})\rho_{cg-\kappa,bg}^{[1]}
&+\shs{\sum_{b_1,\kappa_1}}T_{cb_1,\kappa_1}\rho_{b_1g-\kappa+\kappa_1,bg}^{[2]}(-1)^{N_{b_1}}
+\shs{\sum_{d_1,\kappa_1}}T_{cd_1,\kappa_1}\rho_{d_1g-\kappa-\kappa_1,bg}^{[2]}(-1)^{N_{c}}\\
&-\shs{\sum_{c_1,\kappa_1}}\rho_{cg-\kappa,c_1g-\kappa_1}^{[2]}(-1)^{N_{b}}T_{c_1b,\kappa_1}
-\shs{\sum_{a_1,\kappa_1}}\rho_{cg-\kappa,a_1g+\kappa_1}^{[2]}(-1)^{N_{a_1}}T_{a_1b,\kappa_1},
\end{aligned}
\end{equation}
\begin{equation}\label{rheq2a}
\begin{aligned}
\ii\frac{\pd}{\pd{t}}\rho_{bg-\kappa+\kappa',b'g}^{[2]}
\approx(E_{b}-\leadE{\kappa}+\leadE{\kappa'}-E_{b'})\rho_{bg-\kappa+\kappa',b'g}^{[2]}
&+\shs{\sum_{a_1}}T_{ba_1,\kappa}\rho_{a_1g-\kappa+\kappa'+\kappa,b'g}^{[1]}(-1)^{N_{a_1}}
+\shs{\sum_{c_1}}T_{bc_1,\kappa'}\rho_{c_1g-\kappa+\kappa'-\kappa',b'g}^{[1]}(-1)^{N_{b}}\\
&-\shs{\sum_{c_1}}\rho_{bg-\kappa+\kappa',c_1g-\kappa}^{[1]}(-1)^{N_{b'}}T_{c_1b',\kappa}
-\shs{\sum_{a_1}}\rho_{bg-\kappa+\kappa',a_1g+\kappa'}^{[1]}(-1)^{N_{a_1}}T_{a_1b',\kappa'},
\end{aligned}
\end{equation}
\begin{equation}\label{rheq2b}
\begin{aligned}
\ii\frac{\pd}{\pd{t}}\rho_{dg-\kappa-\kappa',bg}^{[2]}
\approx(E_{d}-\leadE{\kappa}-\leadE{\kappa'}-E_{b})\rho_{dg-\kappa-\kappa',bg}^{[2]}
&+\shs{\sum_{c_1}}T_{dc_1,\kappa}\rho_{c_1g-\kappa-\kappa'+\kappa,bg}^{[1]}(-1)^{N_{c_1}}
+\shs{\sum_{c_1}}T_{dc_1,\kappa'}\rho_{c_1g-\kappa-\kappa'+\kappa',bg}^{[1]}(-1)^{N_{c_1}}\\
&-\shs{\sum_{c_1}}\rho_{dg-\kappa-\kappa',c_1g-\kappa}^{[1]}(-1)^{N_{b}}T_{c_1b,\kappa}
-\shs{\sum_{c_1}}\rho_{dg-\kappa-\kappa',c_1g-\kappa'}^{[1]}(-1)^{N_{b}}T_{c_1b,\kappa'}.
\end{aligned}
\end{equation}
\end{fleqn}
%
%Note that all indices with subscript 1 like $a_1$ , $c_1$ , $\kappa_1$ are summed over.
%Note the summation over all the indices with
Note our convention, that we sum over all indices with subscript 1, e.g., $a_1, \ c_1, \ \kappa_1$. Additionally, phase factors like $(-1)^{N_{b}}$ appear due to order exchange of the lead operators with the dot operators, i.e., $\can_{\kappa}(\ket{b}\otimes\ket{g})=(-1)^{N_{b}}\ket{b}\otimes\can_{\kappa}\ket{g}$.

Summing Eqs.~\eqref{rheq0} and \eqref{rheq1} over all the lead states $\ket{g}$ we get
\begin{fleqn}
\begin{equation}\label{pheq0}
\begin{aligned}
\ii\frac{\pd}{\pd{t}}\Phi_{bb'}^{[0]}=(E_{b}-E_{b'})\Phi_{bb'}^{[0]}
&+\shs{\sum_{a_1,\kappa_1}}T_{ba_1,\kappa_1}\Phi_{a_1b',\kappa_1}^{[1]}
+\shs{\sum_{c_1,\kappa_1}}T_{bc_1,\kappa_1}\Phi_{c_1b',\kappa_1}^{[1]}\\
&-\shs{\sum_{c_1,\kappa_1}}\Phi_{bc_1,\kappa_1}^{[1]}T_{c_1b',\kappa_1}
-\shs{\sum_{a_1,\kappa_1}}\Phi_{ba_1,\kappa_1}^{[1]}T_{a_1b',\kappa_1},
\end{aligned}
\end{equation}
\begin{equation}\label{pheq1}
\begin{aligned}
\ii\frac{\pd}{\pd{t}}\Phi_{cb,\kappa}^{[1]}\approx(E_{c}-\leadE{\kappa}-E_{b})\Phi_{cb,\kappa}^{[1]}
+\shs{\sum_{b_1}}T_{cb_1,\kappa}\Phi_{b_1b}^{[0]}f_{\kappa}
-\shs{\sum_{c_1}}\Phi_{cc_1}^{[0]}T_{c_1b,\kappa}f_{-\kappa},
&+\shs{\sum_{b_1,\kappa_1}}T_{cb_1,\kappa_1}\Phi_{b_1b,-\kappa+\kappa_1}^{[2]}
 +\shs{\sum_{d_1,\kappa_1}}T_{cd_1,\kappa_1}\Phi_{d_1b,-\kappa-\kappa_1}^{[2]}\\
&+\shs{\sum_{c_1,\kappa_1}}\Phi_{cc_1,-\kappa+\kappa_1}^{[2]}T_{c_1b,\kappa_1}
 +\shs{\sum_{a_1,\kappa_1}}\Phi_{ca_1,-\kappa-\kappa_1}^{[2]}T_{a_1b,\kappa_1},
\end{aligned}
\end{equation}
\begin{equation}\label{pheq2a}
\begin{aligned}
\ii\frac{\pd}{\pd{t}}\Phi_{bb',-\kappa+\kappa'}^{[2]}\approx(E_{b}-\leadE{\kappa}+\leadE{\kappa'}-E_{b'})\Phi_{bb',-\kappa+\kappa'}^{[2]}
&-\shs{\sum_{a_1}}T_{ba_1,\kappa}\Phi_{a_1b',\kappa'}^{[1]}f_{\kappa}
 +\shs{\sum_{c_1}}T_{bc_1,\kappa'}\Phi_{c_1b',\kappa}^{[1]}f_{-\kappa'}\\
&-\shs{\sum_{c_1}}\Phi_{bc_1,\kappa'}^{[1]}T_{c_1b',\kappa}f_{-\kappa}
 +\shs{\sum_{a_1}}\Phi_{ba_1,\kappa}^{[1]}T_{a_1b',\kappa'}f_{\kappa'},
\end{aligned}
\end{equation}
\begin{equation}\label{pheq2b}
\begin{aligned}
\ii\frac{\pd}{\pd{t}}\Phi_{db,-\kappa-\kappa'}^{[2]}\approx(E_{d}-\leadE{\kappa}-\leadE{\kappa'}-E_{b})\Phi_{db,-\kappa-\kappa'}^{[2]}
&-\shs{\sum_{c_1}}T_{dc_1,\kappa}\Phi_{c_1b,\kappa'}^{[1]}f_{\kappa}
 +\shs{\sum_{c_1}}T_{dc_1,\kappa'}\Phi_{c_1b,\kappa}^{[1]}f_{\kappa'}\\
&-\shs{\sum_{c_1}}\Phi_{dc_1,\kappa'}^{[1]}T_{c_1b,\kappa}f_{-\kappa}
 +\shs{\sum_{c_1}}\Phi_{dc_1,\kappa}^{[1]}T_{c_1b,\kappa'}f_{-\kappa'},
\end{aligned}
\end{equation}
\end{fleqn}
where we introduced the following notation
\begin{equation}\label{phieqs}
\begin{aligned}
&\Phi_{bb'}^{[0]}=\sum_{g}\rho_{bg,b'g}^{[0]},\quad
\Phi_{cb,\kappa}^{[1]}=\sum_{g}\rho_{cg-\kappa,bg}^{[1]}(-1)^{N_{b}},\quad
\Phi_{bc,\kappa}^{[1]}=\big[\Phi_{cb,\kappa}^{[1]}\big]^{*},\\
&\Phi_{ca,-\kappa-\kappa'}^{[2]}=-\sum_{g}\rho_{cg-\kappa-\kappa',ag}^{[2]},\quad
\Phi_{bb',-\kappa+\kappa'}^{[2]}=+\sum_{g}(1-\delta_{\kappa\kappa'})\rho_{bg-\kappa+\kappa',b'g}^{[2]},\\
&f_{\kappa}\equiv f_{k\leadqn}=(\exp[(\leadE{k}-\mu_{\leadqn})/T_{\leadqn}]+1)^{-1},\quad
f_{-\kappa}\equiv 1-f_{k\leadqn}.
\end{aligned}
\end{equation}
Going from Eq.~\eqref{rheq1} to Eq.~\eqref{pheq1} we also used
\begin{equation}
\begin{aligned}
%\bra{g}\cd_{\kappa}\can_{\kappa}\ket{g}
&\rho_{b_1g-\kappa+\kappa_1,bg}^{[2]}=\delta_{\kappa\kappa_1}\rho_{b_1g-\kappa+\kappa,bg}^{[0]}+(1-\delta_{\kappa\kappa_1})\rho_{b_1g-\kappa+\kappa_1,bg}^{[2]},\\
&\rho_{cg-\kappa,c_1g-\kappa_1}^{[2]}=\delta_{\kappa\kappa_1}\rho_{cg-\kappa,c_1g-\kappa}^{[0]}+(1-\delta_{\kappa\kappa_1})\rho_{cg-\kappa,c_1g-\kappa_1}^{[2]}.
\end{aligned}
\end{equation}
We note that $\Phi_{bb'}^{[0]}$ represents the \textit{reduced density matrix} of the quantum dot. Here we have also assumed that the electrons in the leads are \textit{thermally distributed} according to the Fermi-Dirac distribution $f$ and that this distribution is not affected by the coupling to the quantum dots. This assumption leads to the following relations for Eq.~\eqref{rheq1}
\begin{equation}
\begin{aligned}
&\sum_{g}\rho_{b_1g-\kappa+\kappa,bg}^{[0]}\approx f_{\kappa}\Phi_{b_1b}^{[0]},\quad
\sum_{g}\rho_{cg-\kappa,c_1g-\kappa}^{[0]}\approx f_{-\kappa}\Phi_{cc_1}^{[0]},
\end{aligned}
\end{equation}
for Eq.~\eqref{rheq2a}
\begin{equation}
\begin{aligned}
&\sum_{g}\rho_{a_1g-\kappa+\kappa'+\kappa,b'g}^{[1]}(-1)^{N_{a_1}}\approx-f_{\kappa}\Phi_{a_1b',\kappa'}^{[1]},
&&\sum_{g}\rho_{c_1g-\kappa+\kappa'-\kappa',b'g}^{[1]}(-1)^{N_{b}}\approx f_{-\kappa'}\Phi_{c_1b',\kappa}^{[1]},\\
&\sum_{g}\rho_{bg-\kappa+\kappa',c_1g-\kappa}^{[1]}(-1)^{N_{b'}}\approx f_{-\kappa}\Phi_{bc_1,\kappa'}^{[1]},
&&\sum_{g}\rho_{bg-\kappa+\kappa',a_1g+\kappa'}^{[1]}(-1)^{N_{a_1}}\approx-f_{\kappa'}\Phi_{ba_1,\kappa}^{[1]},\nonumber
\end{aligned}
\end{equation}
and for Eq.~\eqref{rheq2b}
\begin{equation}
\begin{aligned}
&\sum_{g}\rho_{c_1g-\kappa-\kappa'+\kappa,bg}^{[1]}(-1)^{N_{c_1}}\approx-\Phi_{c_1b,\kappa'}^{[1]}f_{\kappa},
&&\sum_{g}\rho_{c_1g-\kappa-\kappa'+\kappa',bg}^{[1]}(-1)^{N_{c_1}}\approx\Phi_{c_1b,\kappa}^{[1]}f_{\kappa'},\\
&\sum_{g}\rho_{dg-\kappa-\kappa',c_1g-\kappa}^{[1]}(-1)^{N_{b}}\approx\Phi_{dc_1,\kappa'}^{[1]}f_{-\kappa},
&&\sum_{g}\rho_{dg-\kappa-\kappa',c_1g-\kappa'}^{[1]}(-1)^{N_{b}}\approx-\Phi_{dc_1,\kappa}^{[1]}f_{-\kappa'}.
\end{aligned}
\end{equation}
Equations \eqref{pheq0}-\eqref{pheq2b} are the central equations for deriving approximate master equation implemented in \texttt{QmeQ}.

\section{\label{App:2vN} Solution of the second-order von Neumann approach equations}

In the \textit{second-order von Neumann} (\emph{2vN}) approach we use the approximations:
\begin{enumerate}
\item[\textbf{1.}] Only terms involving up to two excitations $n=2$ are kept ($\Phi^{[3]}\rightarrow 0$).
\item[\textbf{2.}] A Markov approximation is made to $\Phi^{[2]}$.% which corresponds to setting $\ii\pd_{t}\Phi^{[2]}=0$.
\item[\textbf{3.}] Electrons in the leads are \textit{thermally distributed} according to the Fermi-Dirac distribution $f$.
\end{enumerate}
Pros~(\textbf{+}) and cons~(\textbf{-}) of this approach:
\begin{enumerate}
\item[\textbf{+}] Can describe sequential tunneling, cotunneling, pair-tunneling, and broadening effects \cite{PedersenPRB2005}.
\item[\textbf{+}] Yields exact currents for non-interacting systems with $H_{\mr{Coulomb}}=0$ \cite{KonigPRB1996,KarlstromJPA2013}.
\item[\textbf{-}] With interactions ($H_{\mr{Coulomb}}\neq 0$) the results can only be trusted up to moderate coupling strengths $\Gamma\lesssim T$. Also the temperature $T$ needs to be larger than any Kondo temperature $T_{K}$ in the system \cite{PustilnikJPhysCondensMat2004,PedersenPRB2005,HewsonBook1993}.
\item[\textbf{-}] Can violate the positivity of the reduced density matrix $\Phi^{[0]}$ and does not satisfy the Onsager relations \cite{SejaPRB2016}, when $H_{\mr{Coulomb}}\neq 0$. However, negative occupations are rare compared to the first-order approaches addressed below and occur only for strong couplings. Similarly, the deviations from Onsager's theorem are of third order in the rates ($\Gamma^3$) as shown in \cite{SejaPRB2016}. Thus these features may be used to monitor the validity of results with increasing tunnel coupling.
\item[\textbf{-}] Has long calculation times and large memory consumption compared to the first-order approaches .
\end{enumerate}
After solving the linear inhomogeneous differential equations \eqref{pheq2a} and \eqref{pheq2b} we obtain:
\begin{fleqn}
\begin{equation}
\begin{aligned}
\Phi_{bb',-\kappa+\kappa'}^{[2]}(t)=\frac{1}{\ii}\int_{-\infty}^{t}
\dif{t'}\e^{\ii(\leadE{\kappa}-\leadE{\kappa'}-E_{b}+E_{b'}+\ii\eta)(t-t')}
\Big(&-\shs{\sum_{a_1}}T_{ba_1,\kappa}\Phi_{a_1b',\kappa'}^{[1]}(t')f_{\kappa}
 +\shs{\sum_{c_1}}T_{bc_1,\kappa'}\Phi_{c_1b',\kappa}^{[1]}(t')f_{-\kappa'}\\
&-\shs{\sum_{c_1}}\Phi_{bc_1,\kappa'}^{[1]}(t')T_{c_1b',\kappa}f_{-\kappa}
 +\shs{\sum_{a_1}}\Phi_{ba_1,\kappa}^{[1]}(t')T_{a_1b',\kappa'}f_{\kappa'}
\Big).
\end{aligned}
\end{equation}

\begin{equation}
\begin{aligned}
\Phi_{db,-\kappa-\kappa'}^{[2]}(t)=\frac{1}{\ii}\int_{-\infty}^{t}
\dif{t'}\e^{\ii(\leadE{\kappa}+\leadE{\kappa'}-E_{d}+E_{b}+\ii\eta)(t-t')}
\Big(&-\shs{\sum_{c_1}}T_{dc_1,\kappa}\Phi_{c_1b,\kappa'}^{[1]}(t')f_{\kappa}
 +\shs{\sum_{c_1}}T_{dc_1,\kappa'}\Phi_{c_1b,\kappa}^{[1]}(t')f_{\kappa'}\\
&-\shs{\sum_{c_1}}\Phi_{dc_1,\kappa'}^{[1]}T_{c_1b,\kappa}(t')f_{-\kappa}
 +\shs{\sum_{c_1}}\Phi_{dc_1,\kappa}^{[1]}T_{c_1b,\kappa'}(t')f_{-\kappa'}
\Big).
\end{aligned}
\end{equation}
\end{fleqn}
Here we have added a positive infinitesimal $\eta=+0$ to ensure a proper decay of initial conditions. Here we neglect
the memory by replacing $t'$ by $t$ in the $\Phi^{[1]}$ functions of the integrals (Markov approximation). Then the integration of the exponential factor provides the final result for $\Phi^{[2]}$ functions:
\begin{equation}\label{ssphi2_2vN}
\begin{aligned}
&\Phi_{bb',-\kappa+\kappa'}^{[2]}=
\frac{-\shs{\sum_{a_1}}T_{ba_1,\kappa}\Phi_{a_1b',\kappa'}^{[1]}f_{\kappa}
 +\shs{\sum_{c_1}}T_{bc_1,\kappa'}\Phi_{c_1b',\kappa}^{[1]}f_{-\kappa'}
-\shs{\sum_{c_1}}\Phi_{bc_1,\kappa'}^{[1]}T_{c_1b',\kappa}f_{-\kappa}
 +\shs{\sum_{a_1}}\Phi_{ba_1,\kappa}^{[1]}T_{a_1b',\kappa'}f_{\kappa'}}{\leadE{\kappa}-\leadE{\kappa'}-E_{b}+E_{b'}+\ii\eta},\\
%%%%%%%%%%%%%%%%%%%
&\Phi_{db,-\kappa-\kappa'}^{[2]}=
\frac{-\shs{\sum_{c_1}}T_{dc_1,\kappa}\Phi_{c_1b,\kappa'}^{[1]}f_{\kappa}
 +\shs{\sum_{c_1}}T_{dc_1,\kappa'}\Phi_{c_1b,\kappa}^{[1]}f_{\kappa'}
-\shs{\sum_{c_1}}\Phi_{dc_1,\kappa'}^{[1]}T_{c_1b,\kappa}f_{-\kappa}
 +\shs{\sum_{c_1}}\Phi_{dc_1,\kappa}^{[1]}T_{c_1b,\kappa'}f_{-\kappa'}}{\leadE{\kappa}+\leadE{\kappa'}-E_{d}+E_{b}+\ii\eta}.
\end{aligned}
\end{equation}
We note that the Markov approximation can be made quicker by this simple substitution:
\begin{equation}\label{2vN_Markov_cond}
\begin{aligned}
\ii\left(\frac{\pd}{\pd{t}}+\eta\right)\Phi_{bb',-\kappa+\kappa'}^{[2]}=0,\quad
\ii\left(\frac{\pd}{\pd{t}}+\eta\right)\Phi_{db,-\kappa-\kappa'}^{[2]}=0.
\end{aligned}
\end{equation}
%
%allows us to write $\Phi^{[2]}$ in terms of $\Phi^{[1]}$ as
%
%Here we have added a positive infinitesimal $\eta=+0$ to ensure a proper decay of initial conditions.
The above relations are also expected to hold in the stationary state. After inserting the above expressions into Eq.~\eqref{pheq1}, we obtain the integral equations of the \emph{2vN} method (colors are explained below)
\begin{equation}\label{eq2vN}
\begin{aligned}
%%%%%%%%%%
\ii\frac{\pd}{\pd t}\Phi^{[1]}_{cb,\kappa}&=-(\leadE{\kappa}-E_{c}+E_{b}+\ii\eta)\Phi^{[1]}_{cb,\kappa}
+T_{cb_1,\kappa}f_{\kappa}\Phi^{[0]}_{b_1b}
-\Phi^{[0]}_{cc_1}f_{-\kappa}T_{c_1b,\kappa}\\
%1
&%\phantom{...}
+\frac{T_{cb_1,\kappa_1}\left[
\color{blue}{
T_{b_1c_1,\kappa_1}f_{-\kappa_1}\Phi^{[1]}_{c_1b,\kappa}
+\Phi^{[1]}_{b_1a_1,\kappa}f_{\kappa_1}T_{a_1b,\kappa_1}}
\color{red}{
-T_{b_1a_1,\kappa}f_{\kappa}\Phi^{[1]}_{a_1b,\kappa_1}
-\Phi^{[1]}_{b_1c_1,\kappa_1}f_{-\kappa}T_{c_1b,\kappa}}
\right]}{\leadE{\kappa}-\leadE{\kappa_1}-E_{b_1}+E_{b}+\ii\eta}\\
%2
&%\phantom{...}
+\frac{T_{cd_1,\kappa_1}\left[
\color{blue}{
T_{d_1c_1,\kappa_1}f_{\kappa_1}\Phi^{[1]}_{c_1b,\kappa}
+\Phi^{[1]}_{d_1c_1,\kappa}f_{-\kappa_1}T_{c_1b,\kappa_1}}
\color{red}{
-T_{d_1c_1,\kappa}f_{\kappa}\Phi^{[1]}_{c_1b,\kappa_1}
-\Phi^{[1]}_{d_1c_1,\kappa_1}f_{-\kappa}T_{c_1b,\kappa}}
\right]}{\leadE{\kappa}+\leadE{\kappa_1}-E_{d_1}+E_{b}+\ii\eta}\\
%3
&%\phantom{...}
+\frac{\left[
\color{blue}{
T_{cd_1,\kappa_1}f_{-\kappa_1}\Phi^{[1]}_{d_1c_1,\kappa}
+\Phi^{[1]}_{cb_1,\kappa}f_{\kappa_1}T_{b_1c_1,\kappa_1}}
\color{red}{
-T_{cb_1,\kappa}f_{\kappa}\Phi^{[1]}_{b_1c_1,\kappa_1}
-\Phi^{[1]}_{cd_1,\kappa_1}f_{-\kappa}T_{d_1c_1,\kappa}}
\right]T_{c_1b,\kappa_1}}{\leadE{\kappa}-\leadE{\kappa_1}-E_{c}+E_{c_1}+\ii\eta}\\
%4
&%\phantom{...}
+\frac{\left[
\color{blue}{
T_{cb_1,\kappa_1}f_{\kappa_1}\Phi^{[1]}_{b_1a_1,\kappa}
+\Phi^{[1]}_{cb_1,\kappa}f_{-\kappa_1}T_{b_1a_1,\kappa_1}}
\color{red}{
-T_{cb_1,\kappa}f_{\kappa}\Phi^{[1]}_{b_1a_1,\kappa_1}
-\Phi^{[1]}_{cb_1,\kappa_1}f_{-\kappa}T_{b_1a_1,\kappa}}
\right]T_{a_1b,\kappa_1}}{\leadE{\kappa}+\leadE{\kappa_1}-E_{c}+E_{a_1}+\ii\eta},
%%%%%%%%%%
\end{aligned}
\end{equation}
and
\begin{equation}\label{eq2vN2}
\ii\frac{\pd}{\pd t}\Phi_{bb'}^{[0]}=(E_{b}-E_{b'})\Phi_{bb'}^{[0]}
+\shs{\sum_{a_1,\kappa_1}}T_{ba_1,\kappa_1}\Phi_{a_1b',\kappa_1}^{[1]}
+\shs{\sum_{c_1,\kappa_1}}T_{bc_1,\kappa_1}\Phi_{c_1b',\kappa_1}^{[1]}
-\shs{\sum_{c_1,\kappa_1}}\Phi_{bc_1,\kappa_1}^{[1]}T_{c_1b',\kappa_1}
-\shs{\sum_{a_1,\kappa_1}}\Phi_{ba_1,\kappa_1}^{[1]}T_{a_1b',\kappa_1}.\quad
\end{equation}
Additionally, we impose the normalisation condition for the diagonal reduced density matrix elements:
\begin{equation}
\sum_{b}\Phi_{bb}^{[0]}=1.
\end{equation}
In \texttt{QmeQ} we perform steady-state transport calculations and thus have the additional conditions
\begin{equation}\label{statcond}
\begin{aligned}
\ii\frac{\pd}{\pd{t}}\Phi_{bb'}^{[0]}=0,\quad \text{and}\quad
\ii\frac{\pd}{\pd{t}}\Phi_{cb,\kappa}^{[1]}=0.
\end{aligned}
\end{equation}

Now we describe the numerical procedure implemented in \texttt{QmeQ} for solving the integral equations \eqref{eq2vN} and \eqref{eq2vN2} in the stationary state. This numerical procedure works well for moderate coupling strengths $\Gamma\lesssim T$ and a large bandwidth $D\gg T,\Gamma$. Using Eq.~\eqref{statcond}, the integral equation \eqref{eq2vN} can be cast in the following form
\begin{equation}\label{itereq2vN}
\Phi^{[1]}_{\kappa}=F_{\kappa}+\sum_{\kappa_1}K_{\kappa,\kappa_1}\Phi^{[1]}_{\kappa_1}.
\end{equation}
The function $F_{\kappa}$ corresponds to a solution of Eq.~\eqref{eq2vN} with a local approximation, i.e., terms of the form $\Phi^{[1]}_{ba,\kappa_1}$ which have integrated energy label $\kappa_1$ are neglected (colored in red). We solve Eq.~\eqref{itereq2vN} iteratively with the zeroth iteration given by $\Phi^{[1]}_{\kappa,0}=F_{\kappa}$.
Then the first correction is determined as $\delta\Phi^{[1]}_{\kappa,1}=\sum_{\kappa_1}K_{\kappa,\kappa_1}F_{\kappa_1}$. The higher order corrections are given by $\delta\Phi^{[1]}_{\kappa,n}=\sum_{\kappa_1}K_{\kappa,\kappa_1}\delta\Phi^{[1]}_{\kappa',n-1}$ and the solution is expressed as $\Phi^{[1]}_{\kappa}=\Phi^{[1]}_{\kappa,0}+\sum_{n}\delta\Phi^{[1]}_{\kappa',n}$. In these iterations we have to evaluate Hilbert transforms of the form:
\begin{equation}
\begin{aligned}
H(\Phi^{[1]}_{\kappa})=\frac{1}{\pi}\int_{-D}^{D}\frac{\Phi^{[1]}_{\kappa'}\dif{\leadE{\kappa'}}}{\leadE{\kappa}-\leadE{\kappa'}\pm\ii\eta}
=\frac{1}{\pi}\mc{P}\int_{-D}^{D}\frac{\Phi^{[1]}_{\kappa'}\dif{\leadE{\kappa'}}}{\leadE{\kappa}-\leadE{\kappa'}}\mp\ii\Phi^{[1]}_{\kappa}\theta(D-\abs{\leadE{\kappa}}).
\end{aligned}
\end{equation}
The principal value integrals are efficiently evaluated on equidistant grid with $N$ points with a fast Fourier transform, which has complexity $O(N\log{N})$~\cite{FrederiksenMaster2004,PressBook2007}.

\section{\label{App:1vN}First-order von Neumann approach}
The \textit{first-order von Neumann} ($\emph{1vN}$) approach is obtained with the following approximations to Eqs.~\eqref{pheq0}-\eqref{pheq2b}:
\begin{enumerate}
\item[\textbf{1.}] Only terms involving up to a single excitation $n=1$ are kept ($\Phi^{[2]}\rightarrow 0$).
\item[\textbf{2.}] A Markov approximation is made to $\Phi^{[1]}$.
%\item[\textbf{*.}] Electrons in the leads are \textit{thermally distributed} according to the Fermi-Dirac distribution $f$.
\end{enumerate}
The properties of the approach are:
\begin{enumerate}
\item[\textbf{+}] Can describe sequential tunneling in the presence of coherences.
\item[\textbf{-}] Coupling strength has to be considerably smaller than the temperature $\Gamma \ll T$.
\item[\textbf{-}] Can violate the positivity of the reduced density matrix $\Phi^{[0]}$ \cite{BreuerBook2006} and does not satisfy the Onsager relations \cite{SejaPRB2016,HusseinPRB2014}. Here the same considerations hold as for the \emph{2vN} approach, but problems occur at lower coupling strengths. In particular, the deviation from the Onsager theorem is of second order in the rates ($\Gamma^2$).
\end{enumerate}
We formally integrate Eq.~\eqref{pheq1} (with $\Phi^{[2]}$ neglected) and obtain
\begin{equation}\label{tintphi1}
\begin{aligned}
\Phi_{cb,\kappa}^{[1]}(t)=\frac{1}{\ii}\int_{-\infty}^{t}\dif{t'}\e^{\ii(\leadE{\kappa}-E_{c}+E_{b}+\ii\eta)(t-t')}
\left(T_{cb_1,\kappa}\Phi_{b_1b}^{[0]}(t')f_{\kappa}-\Phi_{cc_1}^{[0]}(t')T_{c_1b,\kappa}f_{-\kappa}\right).
\end{aligned}
\end{equation}
Here we have added a positive infinitesimal $\eta=+0$ to ensure a proper decay of initial conditions. After performing a \textit{Markov} approximation in the above integral, $\Phi_{bb'}^{[0]}(t')\approx\Phi_{bb'}^{[0]}(t)$, and setting $t\rightarrow+\infty$ we get:
\begin{equation}\label{ssphi1_1vN}
\Phi_{cb,\kappa}^{[1]}=
\frac{T_{cb_1,\kappa}\Phi_{b_1b}^{[0]}f_{\kappa}
-\Phi_{cc_1}^{[0]}T_{c_1b,\kappa}f_{-\kappa}}{\leadE{\kappa}-E_{c}+E_{b}+\ii\eta}.
\end{equation}
The same expression \eqref{ssphi1_1vN} is obtained if we use $\ii\pd_{t}\Phi^{[1]}=0$. Combining Eqs.~\eqref{pheq0} and \eqref{ssphi1_1vN} we get the \textit{1vN} approach equations:
\begin{align}\label{ss_1vN}
\begin{aligned}
\ii\frac{\pd}{\pd t}\Phi_{bb'}^{[0]}=(E_{b}-E_{b'})\Phi_{bb'}^{[0]}
+&\sum_{b''\leadqn}\Phi_{bb''}^{[0]}\Big[\sum_{a}\Gamma_{b''a,ab'}^{\leadqn}I_{ba}^{\leadqn-}-\sum_{c}\Gamma_{b''c,cb'}^{\leadqn}I_{cb}^{\leadqn+*}\Big]\\
+&\sum_{b''\leadqn}\Phi_{b''b'}^{[0]}\Big[\sum_{c}\Gamma_{bc,cb''}^{\leadqn}I_{cb'}^{\leadqn+}-\sum_{a}\Gamma_{ba,ab''}^{\leadqn}I_{b'a}^{\leadqn-*}\Big]\\
+&\sum_{aa'\leadqn}\Phi_{aa'}^{[0]}\Gamma_{ba,a'b'}^{\leadqn}[I_{b'a}^{\leadqn+*}-I_{ba'}^{\leadqn+}]
+\sum_{cc'\leadqn}\Phi_{cc'}^{[0]}\Gamma_{bc,c'b'}^{\leadqn}[I_{c'b}^{\leadqn-*}-I_{cb'}^{\leadqn-}],
\end{aligned}
\end{align}
with the normalisation condition $\sum_{b}\Phi_{bb}^{[0]}=1$. In Eq.~\eqref{ss_1vN} the tunneling rate matrix $\Gamma$ is defined as
\begin{equation}\label{mbGam}
\Gamma_{ba,a'b'}^{\leadqn}=2\pi\nu_{F}T_{ba,\leadqn}T_{a'b',\leadqn},
\end{equation}
and the following integral was introduced
\begin{align}\label{ooxi}
&2\pi I_{cb}^{\leadqn\pm}=\int_{-D}^{D}\frac{\dif{E}f\left(\pm \frac{E-\mu_{\leadqn}}{T_{\leadqn}}\right)}{E-E_{cb}+\ii\eta}
=\mc{P}\int_{-D}^{D}\frac{\dif{E}f\left(\pm \frac{E-\mu_{\leadqn}}{T_{\leadqn}}\right)}{E-E_{cb}}
-\ii\pi f(\pm x_{cb}^{\leadqn})\theta(D-\abs{E_{cb}}),\\
\label{ooxi2}
&\quad \text{with}\quad E_{cb}=E_{c}-E_{b},\quad x_{cb}^{\leadqn}=\frac{E_{cb}-\mu_{\leadqn}}{T_{\leadqn}},\quad
f(x)=[\exp(x)+1]^{-1},
\end{align}
which appears after performing the $k$-sums using a flat density of states approximation, i.~e., $\sum_{k}\rightarrow \nu_{F}\int_{-D}^{D} \dif{E}$, with $\nu_{F}$ denoting the density of states at the Fermi level and $2D$ denoting the bandwidth of the leads. For very large bandwidth $D\rightarrow\infty$ compared to the other energy scales the principal part integral in \eqref{ooxi} can be approximated using a digamma function $\Psi$ \cite{AbramowitzBook1972} in the following way
\begin{equation}
\mc{P}\int_{-D}^{D}\frac{\dif{E}f\left(\frac{E-\mu_{\leadqn}}{T_{\leadqn}}\right)}{E-E_{cb}}\stackrel{D\rightarrow\infty}{\approx}
\Real{\Psi\left(\frac{1}{2}+\ii\frac{x_{cb}^{\leadqn}}{2\pi}\right)}-\ln\frac{D}{2\pi T_{\leadqn}}.
\end{equation}
This is the expression used for \pyin{itype=1}.

For the steady state we have $\ii\pd_{t}\Phi_{bb'}^{[0]}=0$ and thus we can determine $\Phi_{bb'}^{[0]}$ from Eq.~\eqref{ss_1vN} and the normalisation condition $\sum_{b}\Phi_{bb}^{[0]}=1$. Lastly, the particle and energy currents can be expressed as
\begin{align}
&I_{\leadqn}=-2\sum_{cb}\Imag\Big[\sum_{b'}\Gamma_{bc,cb'}^{\leadqn}I_{cb}^{\leadqn +}\Phi^{[0]}_{b'b}
-\sum_{c'}\Gamma_{bc,c'b}^{\leadqn}I_{cb}^{\leadqn -}\Phi^{[0]}_{cc'}\Big],\\
&\dot{E}_{\leadqn}=-2\sum_{cb}\Imag\Big[\sum_{b'}\Gamma_{bc,cb'}^{\leadqn}\tilde{I}_{cb}^{\leadqn +}\Phi^{[0]}_{b'b}
-\sum_{c'}\Gamma_{bc,c'b}^{\leadqn}\tilde{I}_{cb}^{\leadqn -}\Phi^{[0]}_{cc'}\Big],
\end{align}
where for the energy currents we introduced the following integral
\begin{align}\label{ooxi_energy}
&\tilde{I}_{cb}^{\leadqn\pm}=\frac{1}{2\pi}\int_{-D}^{D}\frac{\dif{E}Ef\left(\pm \frac{E-\mu_{\leadqn}}{T_{\leadqn}}\right)}{E-E_{cb}+\ii\eta}
=E_{cb}I_{cb}^{\leadqn\pm}+\frac{1}{2\pi}\int_{-D}^{D}\dif{E}f\left(\pm \frac{E-\mu_{\leadqn}}{T_{\leadqn}}\right).
\end{align}
We emphasize that there are many ways to derive the equations referred to here as the \emph{1vN} approach. For example, identical equations result from the real-time diagrammatic technique~\cite{KonigPRL1997,SchoellerEurPhysJSpecTop2009,LeijnsePRB2008} with a perturbation expansion truncated at leading order in the tunnel couplings. In contrast, including next-to-leading order contributions in the real-time diagrammatic technique does not give the same result as the \emph{2vN} approach. Actually, the \emph{2vN} approach includes some terms up to infinite order in the tunnel couplings, which is equivalent to the resonant tunneling approximation \cite{KonigPRB1996} in the diagrammatic approach \cite{KarlstromJPA2013}.

\section{\label{App:Redfield}First-order Redfield approach}
The \textit{first-order Redfield} approach is based on the same approximations as the \textit{1vN} approach to Eqs.~\eqref{pheq0}-\eqref{pheq2b}.
%
%\begin{enumerate}
%\item[\textbf{1.}] Only the terms involving up to single excitation $n=1$ are kept ($\Phi^{[2]}\rightarrow 0$).
%\item[\textbf{2.}] A Markov approximation is made for $\Phi^{[1]}$, but it is different from the one used for the \emph{1vN} approach.
%\item[\textbf{*.}] Electrons in the leads are \textit{thermally distributed} according to the Fermi-Dirac distribution $f$.
%\end{enumerate}
%
Also the \emph{Redfield} approach has the same \textit{pros and cons} as the \emph{1vN} approach. The difference is in the Markov approximation, where we under the integral in Eq.~\eqref{tintphi1} set:
\begin{equation}\label{mark_R}
\Phi_{bb'}^{[0]}(t')\approx \e^{\ii(E_{b}-E_{b'})(t-t')}\Phi_{bb'}^{[0]}(t).
\end{equation}
This gives:
\begin{equation}\label{ssphi1_R}
\begin{aligned}
\Phi_{cb,\kappa}^{[1]}=
\frac{T_{cb_1,\kappa}\Phi_{b_1b}^{[0]}f_{\kappa}}{\leadE{\kappa}-E_{c}+E_{b_1}+\ii\eta}
-\frac{\Phi_{cc_1}^{[0]}T_{c_1b,\kappa}f_{-\kappa}}{\leadE{\kappa}-E_{c_1}+E_{b}+\ii\eta},
\end{aligned}
\end{equation}
Also note that this corresponds to expressing $\Phi^{[0]}$ in the interaction picture under the integral in Eq.~\eqref{tintphi1} and then performing the Markov approximation [$\Phi^{[0]}_{I}(t')\approx \Phi^{[0]}_{I}(t)$]. Using Eqs.~\eqref{pheq0} and \eqref{ssphi1_R} we obtain the \textit{Redfield} approach equations:
\begin{align}\label{ss_RF}
\begin{aligned}
\ii\frac{\pd}{\pd t} \Phi_{bb'}^{[0]}=(E_{b}-E_{b'})\Phi_{bb'}^{[0]}
+&\sum_{b''\leadqn}\Phi_{bb''}^{[0]}\Big[\sum_{a}\Gamma_{b''a,ab'}^{\leadqn}I_{b''a}^{\leadqn-}-\sum_{c}\Gamma_{b''c,cb'}^{\leadqn}I_{cb''}^{\leadqn+*}\Big]\\
+&\sum_{b''\leadqn}\Phi_{b''b'}^{[0]}\Big[\sum_{c}\Gamma_{bc,cb''}^{\leadqn}I_{cb''}^{\leadqn+}-\sum_{a}\Gamma_{ba,ab''}^{\leadqn}I_{b''a}^{\leadqn-*}\Big]\\
+&\sum_{aa'\leadqn}\Phi_{aa'}^{[0]}\,\Gamma_{ba,a'b'}^{\leadqn}[I_{b'a'}^{\leadqn+*}-I_{ba}^{\leadqn+}]
+\sum_{cc'\leadqn}\Phi_{cc'}^{[0]}\,\Gamma_{bc,c'b'}^{\leadqn}[I_{cb}^{\leadqn-*}-I_{c'b'}^{\leadqn-}].
\end{aligned}
\end{align}
%
%Just as in the \textit{1vN} case, we solve Eq.~\eqref{1vN_IntAfterMarkov} under the constraint Eq.~\eqref{rho_sideCondition}.
The definitions of $\Gamma$ and $I$ are the same as in Eqs.~\eqref{mbGam} and \eqref{ooxi}. As a result of the different Markov approximation, both the equation of motion and the currents are different from the \textit{1vN} approach. In \texttt{QmeQ} Eq.~\eqref{ss_RF} is solved for stationary state $\ii\pd_t\Phi_{bb'}^{[0]}=0$ together with normalisation condition $\sum_{b}\Phi_{bb}^{[0]}=1$.

Using Eq.~\eqref{ssphi1_R} the particle and energy currents become
\begin{align}
&I_{\leadqn}=-2\sum_{cb}\Imag\Big[\sum_{b'}\Gamma_{bc,cb'}^{\leadqn}I_{cb'}^{\leadqn +}\Phi^{[0]}_{b'b}
-\sum_{c'}\Gamma_{bc,c'b}^{\leadqn}I_{c'b}^{\leadqn -}\Phi^{[0]}_{cc'}\Big],\\
&\dot{E}_{\leadqn}=-2\sum_{cb}\Imag\Big[\sum_{b'}\Gamma_{bc,cb'}^{\leadqn}\tilde{I}_{cb'}^{\leadqn +}\Phi^{[0]}_{b'b}
-\sum_{c'}\Gamma_{bc,c'b}^{\leadqn}\tilde{I}_{c'b}^{\leadqn -}\Phi^{[0]}_{cc'}\Big].
\end{align}

\section{\label{App:Pauli}Pauli master equation}

The \emph{Pauli} master equation is obtained with the following approximations to Eqs.~\eqref{pheq0}-\eqref{pheq2b}:
\begin{enumerate}
\item[\textbf{1.}] Only the terms involving up to single excitation $n=1$ are kept ($\Phi^{[2]}\rightarrow 0$).
\item[\textbf{2.}] The Markov approximation is made to $\Phi^{[1]}$. %, which corresponds to setting $\ii\pd_{t}\Phi^{[1]}=0$.
\item[\textbf{3.}] The coherences of the reduced density matrix $\Phi^{[0]}$ of the quantum dot are neglected ($\Phi^{[0]}_{bb'}=0$ for $b\neq b'$).
%\item[\textbf{*.}] Electrons in the leads are \textit{thermally distributed} according to the Fermi-Dirac distribution $f$.
\end{enumerate}
The properties of the approach are:
\begin{enumerate}
\item[\textbf{+}] Can describe sequential tunneling.
\item[\textbf{+}] Preserves the positivity of the reduced density matrix $\Phi^{[0]}$ \cite{BreuerBook2006} and satisfies the Onsager relations \cite{AlickiRepMathPhys1976}.
\item[\textbf{-}] The coupling strength has to be considerably smaller than the temperature $\Gamma \ll T$.
\item[\textbf{-}] The energy level splitting $\Delta{E}$ between the states with the same charge has to be considerably smaller than the coupling strength $\Gamma$ \cite{SchultzPRB2009,GoldozianSciRep2016}.
\end{enumerate}
The \emph{Pauli} master equation can be obtained from the \textit{1vN} or the \textit{Redfield} approaches by neglecting the coherences $\Phi_{bb'}^{[0]}$, $b\neq b'$. From Eq.~\eqref{ss_1vN} or \eqref{ss_RF} for the populations $P_{b}=\Phi_{bb}^{[0]}$ we obtain the equations:
\begin{align}\label{paulimeq}
\begin{aligned}
\frac{\pd}{\pd t}P_{b}=&\sum_{a\leadqn}\left[P_{a}\Gamma_{a\rightarrow b}^{\leadqn}f(+x_{ba}^{\leadqn})-P_{b}\Gamma_{b\rightarrow a}^{\leadqn}f(-x_{ba}^{\leadqn})\right]\\
+&\sum_{c\leadqn}\left[P_{c}\Gamma_{c\rightarrow b}^{\leadqn}f(-x_{cb}^{\leadqn})-P_{b}\Gamma_{b\rightarrow c}^{\leadqn}f(+x_{cb}^{\leadqn})\right],
\end{aligned}
\end{align}
where we have denoted $\Gamma_{a\rightarrow b}^{\leadqn}=\Gamma_{ab,ba}^{\leadqn}=\Gamma_{b\rightarrow a}^{\leadqn}=\Gamma_{ba,ab}^{\leadqn}$. In \texttt{QmeQ} Eq.~\eqref{paulimeq} is solved for stationary state $\pd_t P_{b}=0$ together with the normalisation condition $\sum_{b}P_b=1$. Using the populations $P_{b}$ the particle and energy currents are expressed as
\begin{align}
\label{current12}
I_{\leadqn}&=\sum_{ab}[P_{a}\Gamma_{a\rightarrow b}^{\leadqn}f(+x_{ba}^{\leadqn})
-P_{b}\Gamma_{b\rightarrow a}^{\leadqn}f(-x_{ba}^{\leadqn})],\\
\label{heatcurrent12}
\dot{E}_{\leadqn}&=\sum_{bc}\big[P_{b}\Gamma_{b\rightarrow c}^{\leadqn} E_{cb}f(+x_{cb}^{\leadqn})
-P_{c}\Gamma_{c\rightarrow b}^{\leadqn} E_{cb} f(-x_{cb}^{\leadqn})\big],
\end{align}
where $E_{cb}$ and $x_{cb}^{\leadqn}$ are defined in \eqref{ooxi2}.

\section{\label{App:Lindblad}Lindblad equation}
The Lindblad equation \cite{LindbladCMP1976}, which we are using in \texttt{QmeQ}, has the following form:
\begin{align}\label{Lindblad}
\begin{aligned}
\ii\frac{\pd}{\pd t}\Phi_{bb'}^{[0]}=(E_{b}-E_{b'})\Phi_{bb'}^{[0]}
-\frac{1}{2}&\sum_{b''\leadqn}\Phi_{bb''}^{[0]}\Big[\sum_{a}L_{ab''}^{\leadqn*}L_{ab'}^{\leadqn}+\sum_{c}L_{cb''}^{\leadqn*}L_{cb'}^{\leadqn}\Big]\\
-\frac{1}{2}&\sum_{b''\leadqn}\Phi_{b''b'}^{[0]}\Big[\sum_{c}L_{cb}^{\leadqn*}L_{cb''}^{\leadqn}+\sum_{a}L_{ab}^{\leadqn*}L_{ab''}\Big]\\
+&\sum_{aa'\leadqn}\Phi_{aa'}^{[0]}L_{ba}^{\leadqn}L_{b'a'}^{\leadqn*}
+\sum_{cc'\leadqn}\Phi_{cc'}^{[0]}L_{bc}^{\leadqn}L_{b'c'}^{\leadqn*},
\end{aligned}
\end{align}
where the matrix elements of the jump operators are defined as \cite{Kirsanskas2017Lindblad}
\begin{equation}
L_{cb,\leadqn}=\sqrt{2\pi\nu_{F} f(+x_{cb}^{\alpha})}T_{bc,\leadqn},\quad
L_{bc,\leadqn}=\sqrt{2\pi\nu_{F} f(-x_{cb}^{\alpha})}T_{bc,\leadqn}.
\end{equation}
As before we solve Eq.~\eqref{Lindblad} for the stationary state $\ii\pd_t\Phi_{bb'}^{[0]}=0$ supplemented by the normalisation condition $\sum_{b}\Phi_{bb}^{[0]}=1$. Some of the properties of the approach are:
\begin{enumerate}
\item[\textbf{+}] Equation \eqref{Lindblad} is of the first-order type in the rates $\Gamma$ and can describe the sequential tunneling in the presence of coherences.
\item[\textbf{+}] Preserves the positivity of the reduced density matrix $\Phi^{[0]}$ \cite{LindbladCMP1976}.
\item[\textbf{-}] The coupling strength has to be considerably smaller than the temperature $\Gamma \ll T$.
\item[\textbf{-}] Does not satisfy the Onsager relations \cite{Kirsanskas2017Lindblad}. Here, the same considerations hold as for the \emph{2vN} approach, but the deviation from the theorem is of second order in the rates ($\Gamma^2$).
\item[\textbf{-}] Effects, which can arise due to principal part integrals [see Eq.~\eqref{ooxi}] are not included.
\end{enumerate}

\section{\label{App:Which}Which approach to use?}

Having different approaches the question ``Which approach to use?'' can arise. From the pros~(\textbf{+}) and cons~(\textbf{-}) of different approaches it is clear that if a system is small and cotunneling or pair-tunneling is under interest then the \emph{2vN} should be used. On the other hand, the \emph{Pauli} master equation is a reliable choice with low computational cost if only  sequential tunneling is of relevance  and  no coherences between different states develop.

If coherences are important, while tunneling is dominated by sequential events, it is possible to use the \emph{1vN}, \emph{Redfield}, or \emph{Lindblad} approaches. We recommend to use the \emph{Lindblad} approach in the case, when the effects of principal part integrals [see Eq.~\eqref{ooxi}] are not important, because this approach preserves positivity. At the same time we note that the principal parts are required to catch important physics such as renormalization of energy levels in some systems~\cite{WunschPRB2005,PedersenPRB2007,SothmannPRB2010,MisiornyNatPhys2013}. By simulating different systems we also observed that with neglected principal parts the \emph{1vN} and the \emph{Redfield} approaches give the same results.  In Ref.~\cite{SejaPRB2016} the deviation from the Onsager's theorem was found to be slightly larger in the \emph{1vN} approach compared to the \emph{Redfield} approach (with the principal parts included). However, for the stationary state, the density matrix is constant in time within the Schr\"{o}dinger picture. Thus for stationary states it appears to be better to do the Markov limit in the Schr\"{o}dinger picture (the \emph{1vN} approach). So we cannot give a clear-cut answer for what case the \emph{Redfield} or \emph{1vN} approach is better. The clear thing is that for sufficiently weak coupling both approaches do not show substantial differences.

%\bibliographystyle{elsarticle-num}
%\bibliographystyle{apsrev4-1}
%\bibliography{refs_gediminas}

%

\end{document}